\documentclass[fleqn,usenatbib]{mnras}

\usepackage{newtxtext,newtxmath}

\usepackage[T1]{fontenc}

\DeclareRobustCommand{\VAN}[3]{#2}
\let\VANthebibliography\thebibliography
\def\thebibliography{\DeclareRobustCommand{\VAN}[3]{##3}\VANthebibliography}

\usepackage{graphicx}	
\usepackage{amsmath}	
\usepackage{caption}
\newcommand{\myparallel}{{\mkern3mu\vphantom{\perp}\vrule depth 0pt\mkern2mu\vrule depth 0pt\mkern3mu}}

\title[Improved synchrotron energy estimates]{Towards improved synchrotron self absorption energy estimates: accounting for inhomogeneous and non-spherical emitting regions}

\author[F. J. Cowie and R. P. Fender]{
F. J. Cowie$^{1}$\thanks{E-mail: fraser.cowie@physics.ox.ac.uk} and
R. P. Fender$^{1,2}$
\\
$^{1}$Department of Physics, University of Oxford, Denys Wilkinson Building, Keble Road, Oxford OX1 3RH, UK \\
$^{2}$Department of Astronomy, University of Cape Town, Private Bag X3, 7701 Rondebosch, South Africa\\
}

\date{Accepted XXX. Received YYY; in original form ZZZ}

\pubyear{2025}

\begin{document}
\label{firstpage}
\pagerange{\pageref{firstpage}--\pageref{lastpage}}
\maketitle

\begin{abstract}
Synchrotron self absorption (SSA) is seen across a variety of astrophysical sources, and observation of an SSA peak in the spectrum is a powerful tool for estimating the physical conditions and the minimum energy of the emitting region. We begin with the (re)derivation of the usual SSA parameter estimates, carefully considering dependencies and assumptions, obtaining the most accurate traditional SSA minimum energy equations currently available. Traditional methods rely on the assumption that the emitting region is quasi-spherical and homogeneous. However, many observations of SSA show that the spectral index at frequencies below the peak is less than the expected $+2.5$ (non-thermal) or $+2$ (thermal). We argue that an inhomogeneous emitting region is the most likely explanation in many cases. Power law inhomogeneous cylindrical slab and broken power law inhomogeneous sphere models are used to investigate how the presence of inhomogeneity affects parameter estimates using traditional SSA methods. We find that in some cases inhomogeneity can lead to traditional SSA methods underestimating the minimum energy and the size of the emitting region by over an order of magnitude. Quantitative correction factors are found which can be applied to traditional estimates to correct for inhomogeneity, depending on the value of the observed flattened spectral index and the range in frequency over which this value is observed. Furthermore, we derive simple correction factors for non-spherical homogeneous emitting regions. Finally, we explore the effects of inhomogeneity on measurements of polarisation around the spectral peak, and on lightcurves for expanding emitting regions.


\end{abstract}




\section{Introduction}\label{sec:intro}

The observation of synchrotron emission is extensively used to estimate the energetics of astrophysical phenomena. Measurement of the radio synchrotron luminosity probes the energy contained within ultra-relativistic particles and the magnetic field these particles reside in, which together make up a synchrotron emitting plasma. This energy can then be used to estimate the kinetic energy of the phenomenon responsible for the particle acceleration leading to the synchrotron radiating ultra-relativistic particles. This makes energetic estimates from measurements of synchrotron radiation invaluable in understanding the physics of a variety of astrophysical sources.


It has been long understood that given the luminosity and volume of a synchrotron emitting plasma, that the minimum energy required to produce this plasma can be calculated \citep{burbidge_1956}. For a given luminosity and volume, a minimum energy exists close to energy equipartition between the non-thermal particles and the magnetic field. Significant deviations from equipartition require total energies which are orders of magnitude greater in order to reproduce observations.


For the case of spatially resolved astrophysical sources, such as supernovae remnants or large scale active galactic nuclei (AGN) jets the emitting volume, required for the energy calculation, can be easily estimated. This is more complicated for unresolved synchrotron sources, although several methods can be used (see  Section~1 of \citealt{fender_bright_2019} for a summary). The most accurate and widely used method for unresolved sources is instead to use the observation of the synchrotron self absorption (SSA) peak frequency and amplitude to introduce an additional relationship between the size of the emitting region and the magnetic field. This allows for the minimum energy as a function of the size of the emitting region to be found, rather than as a function of the magnetic field as in the known size case. This minimum energy is close (within a factor of unity) to the equipartition estimate if the size was known independently. This method is developed and discussed in \cite{scott_1977}, \cite{bnp_2013}, \cite{fender_bright_2019}, and \cite{matsumoto_2023} and requires the SSA peak in the synchrotron spectrum to be observed, which is often done at radio wavelengths.


All these calculations provide a robust minimum energy estimate and an associated size and magnetic field, as most (but not all) deviations from the basic set of assumptions used, such as equipartition, result in increases to the energy. These assumptions are discussed in detail in Section~\ref{sec:motivation}. 


However, one key assumption made in the application of SSA energy estimates is homogeneity of the emitting region. It is not clear how violating this assumption effects these traditional SSA estimates, and exploring this is the primary topic of this work. If the synchrotron emitting region is homogeneous in optical depth and specific intensity (and therefore necessarily homogeneous in  magnetic field strength and emitting particle density) then at frequencies lower than the SSA peak the spectral index (defined as $F_\nu \propto \nu^\alpha$) will be the expected $+2.5$ (or $+2$ if the spectrum is dominated by electrons of a single temperature) \citep{pacholczyk}. This is known as the optically thick or self absorbed region of the spectrum. Inhomogeneous synchrotron sources were first studied in an attempt to explain extragalactic flat spectrum ($\alpha \sim 0$) radio sources. Numerous authors have shown that introducing inhomogeneity \textit{changes the spectral index at frequencies lower than the spectral peak} e.g. \cite{ozernoy_1971}, \cite{condon_1973}. Authors introduced inhomogeneity using a variety of different models, such as axially symmetric power law slabs \citep{de_bruyn_1976} and in spherical geometries (\citealt{gould_1979}, \citealt{band_grindlay_1985}). A satisfactory explanation for the flat spectra of some extragalactic radio sources was eventually found using the inhomogeneous synchrotron model of an isothermal conical jet by \cite{blandford_konigl_1979}. 

However, there are many other astrophysical sources of varying nature which show spectral indices below the spectral peak which are significantly flatter than the expected $+2.5$ but still $>0$, and in general are not explained by the model presented in \cite{blandford_konigl_1979}. Many examples of these sources with flatter spectral indices below the spectral peak are some form of synchrotron transient where the peak in the spectrum is associated with the peak of the lightcurve as the source undergoes the transition from optically thick to thin synchrotron emission. This is often evident from a change in the measured spectral index from positive to negative through the peak of the lightcurve. Examples of this behaviour are seen in radio flares from gamma ray bursts (GRBs) (e.g. \citealt{laskar_2013}, \citealt{laskar_2019}), tidal disruption events (TDEs) (e.g. \citealt{stein_2021}, \citealt{andreoni_2022}, \citealt{goodwin_2022}), fast rising blue optical transients (FBOTs) (e.g. \citealt{margutti_2019}, \citealt{ho_2019}, \citeyear{ho_2022}, \citealt{nayana_2021}, \citealt{nayana_2025}), and supernovae (SNe) (e.g. \citealt{sfaradi_2024}). It is perhaps most often observed in the radio flares from X-ray binaries (XRBs) during state transitions e.g. \cite{fender_2023} where 85 optically thick to thin flare peaks, none of which show spectral indices above $+1.5$ (their Figure~9). It is in all of these sources that traditional SSA energy estimates, assuming a homogeneous source, are widely applied in order to extract key physics such as the minimum energy. 



We note that not all radio flares represent the transition of an expanding plasma from optically thick to thin. For these optically thin flares, thought to be caused due to extended periods of particle acceleration \citep{fender_bright_2019}, SSA energy estimates can only provide limits on the minimum energy and other parameters. This is discussed further in Section~\ref{sec:homog}. In this work, we consider peaks in the spectrum of synchrotron emission which can be identified as SSA as these are also where the traditional methods are valid. We also note that other effects can mimic the positive spectral index of synchrotron self absorption, such as free-free absorption \citep{fender_2001}. However, these scenarios can often be differentiated using variability or other arguments \citep{chevalier_1998}.


Numerous authors have noted the flattening of the optically thick spectral index and while several explanations have been put forward, few have been investigated in detail. These various scenarios are discussed and evaluated in Section~\ref{sec:motivation}. If inhomogeneity is a likely cause of these flattened spectral indices, then energy and size estimates using SSA, where homogeneity is a key assumption may require significant correction. \cite{marscher_1977} explored the effects of inhomogeneity on derived physical parameters for sources where Very Long Baseline Interferometry (VLBI) measurements can be used to help constrain the size. However, in the majority of cases, in particular when studying flares from time-variable synchrotron sources, sources cannot be resolved.

In this work we use two simple inhomogeneous emitting region models to explore the impact of inhomogeneity and deviations from sphericity on traditional SSA estimates. These models are a face-on power law inhomogeneous cylindrical slab, and a radial power law inhomogeneous sphere. We derive correction factors on measured energies, sizes and magnetic fields based on observable features, and discuss the implications of these. Furthermore, using the slab model we explore how inhomogeneity impacts measured polarisation properties, and the predicted time evolution of expanding sources. Section~\ref{sec:homog} carefully (re)derives the traditional (homogeneous) SSA estimate equations, resulting in the most accurate set of traditional SSA estimate equations available to date. Section~\ref{sec:motivation} explores the different explanations for a flattening of the spectral index below the spectral peak, and arrives at the conclusion that inhomogeneity is a likely candidate in most cases. Section~\ref{sec:model} outlines the models used throughout this work. Section~\ref{sec:results} explores the energetic, polarisation and time evolution implications of the models when compared to a homogeneous source. Section~\ref{sec:discussion} discusses the broad implications of inhomogeneity in synchrotron sources and Section~\ref{sec:conclusion} summarises our findings. Finally, the Appendices~\ref{sec:full_equations}-\ref{sec:doppler_and_redshift} contain detailed traditional SSA equations with various correction factors and their derivation.

Throughout this work all equations and values use cgs units and the spectral index, $\alpha$, is defined such that $F_\nu \propto \nu^\alpha$.

\section{Homogeneous synchrotron self absorption energy estimates}\label{sec:homog}


The model commonly applied to observations which forms the basis of the traditional SSA energy estimation methods such as those presented in e.g. \citealt{fender_bright_2019}, \citealt{matsumoto_2023}, is that of an isolated, stationary, homogeneous, non-thermal synchrotron emitting plasma. In the following we summarise the derivation of the flux density spectrum of this model as presented in \cite{pacholczyk}, while clearly stating the various approximations used. We then present accurate minimum energy, size and magnetic field formula, with full dependence on the energy distribution of the synchrotron emitting electrons.

We start with the energy distribution of the synchrotron emitting electron population (the only electron population assumed to be present) in the plasma, which is assumed to follow single unbounded power law:

\begin{equation}\label{eq:N0_definition}
    N(E) = N_0 E^{-p} , 
\end{equation}

\noindent where $N(E)$ is the density-density of electrons (i.e. $N(E)dE$ gives the number density of electrons in the energy range between $E$ and $E+dE$), $E$ is the electron energy, and $N_0$ is a normalisation with units depending on $p$, the electron energy index. We note that in other works $N$ is often the total number of electrons in the plasma, however our definition is more convenient for when inhomogeneity is discussed later on in Section~\ref{sec:model}. The fiducial case of $p=2$ is used throughout this work, but results unless otherwise stated are only weakly dependent on $p$. For this case of $p=2$, $N_0$ is an energy density (units of energy per volume). Throughout this work we assume that the synchrotron emitting electrons are ultra-relativistic ($\gamma >> 1$), for the general case of synchrotron emission and where this approximation fails see e.g. \cite{sokolov_synchrotron_1968}. 

The synchrotron emission from a single ultrarelativistic electron in a uniform magnetic field is characterised by the critical frequency, $\nu_{cr}$, above which the synchrotron radiation exponentially decreases and below which the specific intensity is $\propto \nu^\frac{1}{3}$. The critical frequency is given by:

\begin{equation}\label{eq:nu_crit}
    \nu_{cr}(E,B,\alpha_p) = \frac{3}{4 \pi} \sin{\alpha_{p}} \gamma^2 \frac{e B}{m_e c} = c_1 B \sin{\alpha_{p}} E^2 ,
\end{equation}

\noindent where $\gamma$ is the Lorentz factor of the electron, related to the electron energy by $E=\gamma m_e c^2$, $m_e$ is the electron mass and $c$ is the speed of light. The magnetic field amplitude in the emitting volume is $B$, and the pitch angle of the electron (the angle which the electron velocity vector makes with the magnetic field vector) is $\alpha_p$. The various constants can be combined into the constant $c_1$, introduced in \cite{pacholczyk}, and given in Appendix~\ref{sec:full_equations}. Following \cite{pacholczyk} the properties of a single synchrotron emitting electron are used to calculate the synchrotron emission and absorption coefficients from the non-thermal distribution of electrons in a uniform magnetic field, with an isotropic distribution of pitch angles, and making the assumption that $\alpha_p = \theta$, where $\theta$ is the angle between the magnetic field direction and the line of sight of the observer. This is valid due to the beaming of radiation from the ultra-relativistic electrons, meaning that the only emission observed is from electrons which have an instantaneous velocity almost towards the observer at some point on their "orbit" around the magnetic field lines.

The emission and absorption coefficients are calculated in \cite{pacholczyk} and depend on the polarisation of the emission relative to the projected magnetic field in the plane perpendicular to the direction of emission (the plane of the sky). We can split the emission into the fractions perpendicular and parallel to the projected magnetic field, and the resulting emission and absorption coefficients are given by:

\begin{equation}\label{emmission_eq}
    \epsilon_\nu^{\perp , \myparallel} = \frac{1}{2} c_5(p) N_0 (B \sin{\theta})^\frac{p+1}{2} \left(\frac{\nu}{2c_1}\right)^\frac{1-p}{2} \left(1 \pm \frac{3p+3}{3p+7}\right)
\end{equation}

\begin{equation}\label{absorption_eq}
    \kappa_\nu^{\perp , \myparallel} = c_6(p) N_0 (B \sin{\theta})^\frac{p+2}{2} \left(\frac{\nu}{2c_1}\right)^\frac{-(p+4)}{2} \left(1 \pm \frac{3p+6}{3p+10}\right)
\end{equation}

\noindent where $c_5$ and $c_6$ are slowly varying functions of $p$ given in \cite{pacholczyk} and Appendix~\ref{sec:full_equations}. Then the total emission coefficient, $\epsilon_\nu = \epsilon_\nu^{\perp} + \epsilon_\nu^{\myparallel}$, and average absorption coefficient, $\kappa_\nu = \frac{ \kappa_\nu^{\perp} +  \kappa_\nu^{\myparallel}}{2}$, are used in the solution to the radiative transfer equation. We assume the emission and absorption coefficients are constant along the line of sight in the source, and no incident radiation is present. This is the slab geometry approximation. The total specific intensity is then:

\begin{equation}\label{intensity_eq}
    I_\nu = \frac{\epsilon_\nu}{\kappa_\nu} (1 - e^{-\kappa_\nu s}) , 
\end{equation}

\noindent where $s$ is the depth of the source along the line of sight. We note that the averaging of the absorption coefficients implies that the component of the magnetic field in the plane of the sky is randomly variable, while the component of the magnetic field along the line of sight is uniform with magnitude $B\cos\theta$. For a truly uniform magnetic field, the specific intensity for each polarisation should be calculated separately, $I^\perp_\nu$ and $I^\myparallel_\nu$, and added together. However, this only leads to a maximum of a 10\% deviation in the spectrum and so is a minor effect. Throughout this work, we take $\theta=\frac{\pi}{2}$, so that the magnetic field has no component along the line of sight, and from the averaging of the absorption coefficients is assumed to be tangled in the plane of the sky. Adopting a truly random 3D isotropic magnetic field results in fluxes in the optically thin and thick regime of the spectrum which are altered by $\sim20\%$ from the $\theta=\frac{\pi}{2}$ and a slightly ($\sim10\%$) broadened SSA peak. These effects (on the spectrum and resulting energy estimates etc.) are sub-dominant when compared to the effects of inhomogeneity discussed later and so can be safely neglected.

When this model is applied to observations a face on cylindrical slab geometry with depth $s$ and radius $R$ is typically assumed, and therefore the flux at the observer is:

\begin{equation}\label{flux_eq}
    F_\nu = \Omega I_\nu = \pi \frac{R^2}{D^2} I_\nu , 
\end{equation}

\noindent where $\Omega$ is the solid angle of the source on the sky, $D$ is the distance to the source. It is important to note this is \textit{not} the synchrotron spectrum of a homogeneous synchrotron emitting sphere with radius $R_s$. If $\pi R^2 s = \frac{4}{3} \pi R_s^3$ and $R=R_s$ (matching the volumes and the observed solid angle), then the spectrum from the homogeneous cylindrical slab is very similar to the spectrum of a homogeneous emitting sphere, with a correction factor of $<5\%$ for $p=2$ (see \citealt{band_grindlay_1985} for more details). In particular if the above holds the optically thin and thick limits of the spectra will be identical in the sphere and cylindrical cases. In this case we refer to the slab as being quasi-spherical. This slab approach was used for example to model emission from supernova by \citep{chevalier_1998}. For the rest of this Section we will assume the model is quasi-spherical, and the implications of relaxing this assumption are discussed in Section~\ref{sec:results_spehrical}.

\begin{figure}
 \includegraphics[width=\columnwidth]{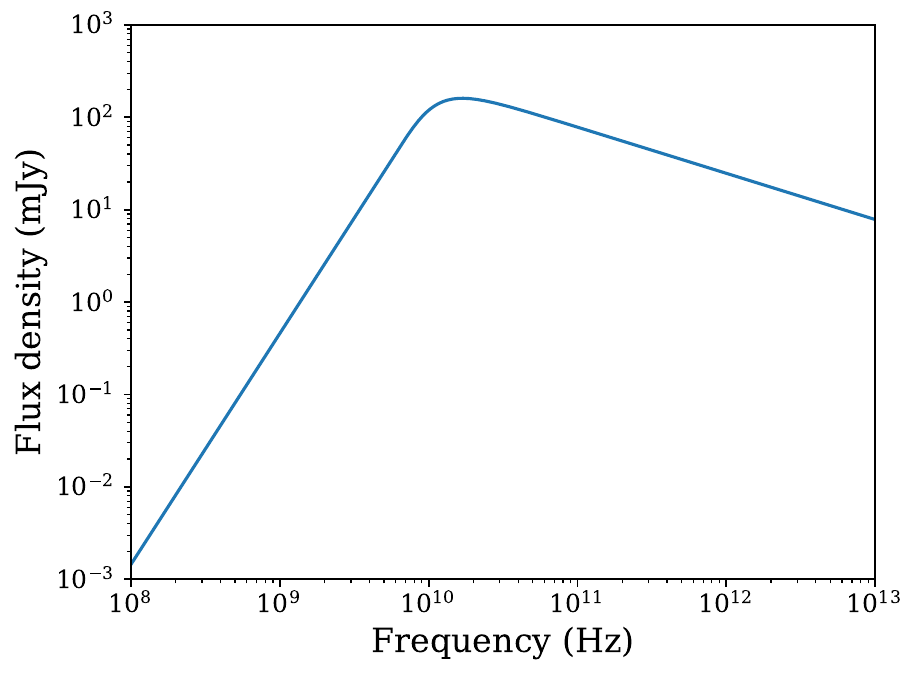}
 \caption{Idealised synchrotron spectrum for the homogeneous cylindrical slab model presented in Section~\ref{sec:motivation} and in \protect\cite{pacholczyk}. This is the model commonly fit to observations, and the model assumed when using SSA methods to measure physical parameters of the emitting region. This synchrotron spectrum has $p=2$.}
 \label{fig:default_sync_spec}
\end{figure}

Using equation~\eqref{flux_eq}, these calculations result in the spectrum shown in Figure~\ref{fig:default_sync_spec}. The spectrum has several notable observable features. The peak of the spectrum, due to SSA, occurs at a frequency of $\nu_\text{peak}$ where the optical depth is $\tau_\text{max}$. This is related to the frequency at which the optical depth of the emitting region is unity, $\nu_1$ by:

\begin{equation}\label{numax_and_nu1}
    \nu_\text{peak} = \frac{\nu_1}{\tau_\text{peak}^{\frac{2}{p+4}}} ,
\end{equation}

\noindent where $\tau_\text{peak}$ is given by\footnote{In the case where a peak in a lightcurve rather than a spectrum is observed, this relationship can differ slightly and becomes dependent on the specific model of time evolution assumed, see e.g. \cite{vdL} and \cite{tetarenko_2017}.}:

\begin{equation}\label{taumax}
    e^{\tau_\text{peak}} = 1+\frac{p+4}{5}\tau_\text{peak} .
\end{equation}

\noindent The frequency at which the optical depth of the cylindrical slab emitting region is unity, $\nu_1$, is given by:

\begin{equation}\label{nu1_eqn}
    \nu_1 = 2 c_1 (s c_6(p))^\frac{2}{p+4} N_0^\frac{2}{p+4} (B \sin{\theta})^\frac{p+2}{p+4} .
\end{equation}

\noindent The flux at this frequency is given by equations \eqref{emmission_eq}, \eqref{absorption_eq}, \eqref{intensity_eq}, and \eqref{flux_eq} for $\nu=\nu_1$:

\begin{equation}\label{fnu1_eqn}
    F_{\nu_1} = \Omega \frac{c_5(p)}{c_6(p)} (B \sin{\theta})^{-\frac{1}{2}} \left(\frac{\nu_1}{2 c_1}\right)^{\frac{5}{2}} (1-e^{-1}) .
\end{equation}

\noindent This is related to the flux at the measured spectral peak by a factor depending only on $p$ \citep{pacholczyk}. Finally, the spectral index when $\nu >> \nu_1$, at frequencies where the region is optically thin, is given by:

\begin{equation}\label{alpha_thin}
    \alpha_{\text{thin}} = \frac{1-p}{2} ,
\end{equation}

\noindent whereas for $\nu << \nu_1$, at frequencies where the region is optically thick, the spectral index is $\alpha_\text{thick} = +2.5$, independent of any model parameters.

\subsection{Traditional SSA estimates}\label{sec:trad_estimates}

We can now use the above to derive the minimum energy required to produce a given synchrotron spectrum within this model, following a similar procedure to \cite{fender_bright_2019}, but with two key differences. First, we maintain the full dependence on $p$ throughout, and second, we do not use the integrated synchrotron luminosity of the source as a proxy for the details of the electron energy distribution. Instead we work with the electron energy distribution directly (e.g. equation 3 in \citealt{fender_bright_2019} is replaced with our equation \ref{eq:e_energy_nu}). 

The electron energy distribution is now assumed to be bounded by a maximum and minimum cut-off energy, $E_\text{max}$ and $E_\text{min}$ respectively, rather than unbounded as was used to generate the spectrum of this model. The effect on the spectrum of relaxing this unbounded assumption results in observable frequency breaks at $\nu_\text{min}=\nu_\text{cr}(E_\text{min})$ and $\nu_\text{max}=\nu_\text{cr}(E_\text{max})$. The details of these breaks are discussed further in Section~\ref{sec:motivation}. For a given spectrum the energy in the magnetic field and in the ultra-relativistic electrons are given respectively by:

\begin{equation}\label{mag_energy}
    E_B = \frac{B^2}{8 \pi} V , 
\end{equation}
\begin{align}
    E_e &= V \int^{E_\text{max}}_{E_\text{min}} N(E) E dE \nonumber \\ &= VN_0 \int^{E_\text{max}}_{E_\text{min}} E^{1-p} dE = V N_0 K_E \\ &=V N_0 B^\frac{p-2}{2}\int^{\nu_\text{max}}_{\nu_\text{min}} \frac{1}{2} \nu^{-\frac{p}{2}} c_1^\frac{p-2}{2} d\nu = V N_0 B^\frac{p-2}{2} K_\nu . \label{eq:e_energy_nu}
\end{align}

\noindent Here we have chosen two ways to parametrise the energy in the electrons depending on whether details of the electron energy spectrum are known, and therefore $K_E = \int^{E_\text{max}}_{E_\text{min}} E^{1-p} dE$ can be calculated; or whether observational constraints on the break frequencies are known, so $K_\nu = \int^{\nu_\text{max}}_{\nu_\text{min}} \frac{\nu^{-\frac{p}{2}}}{2} c_1^\frac{p-2}{2} d\nu$ can be calculated. These two parametrisations are therefore useful in different scenarios. Here we present equations which use $K_\nu$ (hereafter the frequency cut-off formalism) as these may be more useful in observational cases, but the derivation and full set of equations for the $K_E$ case (hereafter the energy cut-off formalism) are presented in Appendix~\ref{sec:energy_cutoff_form}. 

In practice, it can be difficult to accurately constrain $E_\text{min}$ and $E_\text{max}$ or $\nu_\text{min}$ and $\nu_\text{max}$ from observations. In practice the frequency cut-off formalism is often used and $\nu_\text{min}$ and $\nu_\text{max}$ are taken to be the minimum and maximum observed frequencies of synchrotron emission. However, we caution that in many cases, this is equivalent to assuming a narrow, unphysical range of electron energies. This can result in $K_{\nu}$ being significantly underestimated despite the weak dependence on the cut-offs. We therefore provide the full dependence on the electron energy distribution in the two formalisms to encourage readers as far as possible to consider the true uncertainty on frequency or energy cut-offs and propagate this through to physical estimates, for example using a Monte Carlo framework. Furthermore, the energy cut-off formalism can be used in some observational contexts when the SSA peak is observed, by constraining $E_\text{min}$ using equation~\eqref{eqn:gamma_min_constraint}, and constraining $E_\text{max}$ from physical arguments.

Given that the model is quasi-spherical, we know $V\propto R^3$. From equation~\eqref{fnu1_eqn} we have for a constant $F_{\nu_1}$ and $\nu_1$ that $B\propto R^4$. Similarly, using this in equation~\eqref{nu1_eqn} we find that $N_0\propto R^{-(2p+5)}$. Therefore, for a given observed $F_{\nu_1}$ and $\nu_1$, we have a strong dependence (as noted by e.g. \citealt{fender_bright_2019}) on the unknown size for both the magnetic and electron energies:

\begin{equation}\label{eq:enrgy_dependence_size}
    E_\text{total} = E_B+E_e \propto R^{11} + R^{-6} . 
\end{equation}

\noindent The total energy required to generate the spectrum can then be minimised with respect to $R$, and we find that the minimum energy required to generate a given synchrotron spectrum in this model is when:

\begin{equation}\label{eq_min_eng_condition}
    E_B = \frac{6}{11} E_e . 
\end{equation}

\noindent This is close to equipartition ($E_B=E_e$), and close to the minimum energy condition for an observed synchrotron emitting region of known volume ($E_B = \frac{3}{4} E_e$, as in e.g. \citealt{burbidge_1956}, \citealt{cowie_2026}). We can now solve equations \eqref{nu1_eqn}, \eqref{fnu1_eqn} and \eqref{eq_min_eng_condition}, under the assumption of quasi-sphericity, $s=\frac{4}{3} R$, in order to obtain a formula for the minimum energy required to produce a given synchrotron spectrum from a region of unknown size:

\begin{equation}\label{eq:energy_homog_nu_form}
    E_\text{homog, min} = \frac{17}{6} \frac{1}{8\pi} \frac{4\pi}{3} k_1^{-\frac{10}{17}} k_{3\text{,}\nu}^{11} F_{\nu_1}^\frac{20}{17} D^\frac{40}{17} \nu_1^\frac{11p-56}{34} K_\nu^\frac{11}{17} .
\end{equation}

\noindent We also obtain equations for the size and magnetic field this minimum energy occurs at:

\begin{equation}\label{eq:r_homog_nu_form}
    R_\text{homog, eq} = k_1^{-\frac{4}{17}} k_{3\text{,}\nu} F_{\nu_1}^\frac{8}{17} D^\frac{16}{17} \nu_1^\frac{p-36}{34} K_\nu^\frac{1}{17} ,
\end{equation}

\begin{equation}\label{eq:B_homog_nu_form}
    B_\text{homog, eq} = k_1^\frac{1}{17} k_{3\text{,}\nu}^4 F_{\nu_1}^{-\frac{2}{17}} D^{-\frac{4}{17}} \nu_1^\frac{2p+13}{17} K_\nu^\frac{4}{17}, 
\end{equation}

\noindent where $k_1$ and $k_{3,\nu}$ are values which are weakly dependent on $p$ given in Appendix~\ref{sec:full_equations}. We leave the formulae in cgs units, with the $p$ dependence, and fairly unsimplified, so that their derivation is clearer and so that they may be used generally. Furthermore, in Appendix~\ref{sec:full_equations} we present versions of the above formulae considering corrections for bulk relativistic motion, cosmological redshift, the presence of non-thermal protons and deviations from equipartition. We encourage readers to use these more detailed equations to fully characterise the uncertainty on parameters calculated using SSA techniques.

In order to be able to effectively apply the above formula to an observed spectrum, the parameters $p$, $\nu_1$ and $F_{\nu_1}$ must be calculated from observed parameters. $p$ can be determined from $\alpha_\text{thin}$ using equation~\eqref{alpha_thin}. $\nu_1$ is calculated using equations~\eqref{numax_and_nu1} and \eqref{taumax}. Finally $F_{\nu_1}$ can be related to the flux at $\nu_\text{peak}$, $F_{\nu_\text{peak}}$, by combining equations~\eqref{flux_eq} and \eqref{fnu1_eqn} to obtain:

\begin{equation}\label{eq:fnu1_and_fpeak}
    F_{\nu_1} = F_{\nu_\text{peak}} \frac{1-e^{-1}}{(\frac{\nu_\text{peak}}{\nu_1})^{\frac{5}{2}} \left(1-e^{ \left( -\left(\frac{\nu_\text{peak}}{\nu_1}\right)^\frac{-(p+4)
}{2} \right) }\right)} .
\end{equation}

\noindent This entire process from observables to physical parameters is provided in clear modular functions in our public code \footnote{\url{https://github.com/Fraserjcowie/SSA_estimates}}. For example allowing a consistent estimation of energies across a population of sources (Cowie et al. in prep.). The clear dependencies of each physical parameter allows for easy understanding of how large uncertainties in observed parameters (for example $\nu_\text{min}$ and $\nu_\text{max}$) propagate through to the physical parameter estimations. In this case, estimation of the uncertainty lends itself well to a Monte Carlo sampling, as has already been performed in optically thin, known volume synchrotron case (e.g. \citealt{cowie_2026}). 

We can also recast the above formulae into a more easy to use form for a specific value of $p$. For the $p=2$ case \textbf{only} the formulae become:

\begin{align}
    E_\text{homog, min} = & 4.37\times10^{35}~\text{erg} \cdot \left(\frac{F_{\nu_\text{peak}}}{1~\text{mJy}}\right)^{\frac{20}{17}} \cdot \left(\frac{D}{1~\text{kpc}}\right)^{\frac{40}{17}} \cdot \nonumber \\ &\left(\frac{\nu_\text{peak}}{1~\text{GHz}}\right)^{-1} \cdot \left(\ln{\left(\frac{\nu_\text{max}}{\nu_\text{min}}\right)}\right)^{\frac{11}{17}} ,
\end{align}

\begin{align}
    R_\text{homog, eq} = & 2.40\times10^{12}~\text{cm} \cdot \left(\frac{F_{\nu_\text{peak}}}{1~\text{mJy}}\right)^{\frac{8}{17}} \cdot \left(\frac{D}{1~\text{kpc}}\right)^{\frac{16}{17}} \cdot \nonumber \\ &\left(\frac{\nu_\text{peak}}{1~\text{GHz}}\right)^{-1} \cdot \left(\ln{\left(\frac{\nu_\text{max}}{\nu_\text{min}}\right)}\right)^{\frac{1}{17}} ,
\end{align}

\begin{align}
    B_\text{homog, eq} = & 2.58\times10^{-1}~\text{G} \cdot \left(\frac{F_{\nu_\text{peak}}}{1~\text{mJy}}\right)^{-\frac{2}{17}} \cdot \left(\frac{D}{1~\text{kpc}}\right)^{-\frac{4}{17}} \cdot \nonumber \\ &\left(\frac{\nu_\text{peak}}{1~\text{GHz}}\right) \cdot \left(\ln{\left(\frac{\nu_\text{max}}{\nu_\text{min}}\right)}\right)^{\frac{4}{17}} .
\end{align}

\noindent We stress the above formula should only be used in the $p=2$ case, and for accurate determinations of the parameters where $p$ has an associated uncertainty, or is known to be $\neq2$, equations \eqref{eq:energy_homog_nu_form} to \eqref{eq:B_homog_nu_form} should be used (or ideally the equations with all correction factors included in Appendix~\ref{sec:full_equations}).

All of the above formula give the minimum energy required to produce the observed synchrotron spectrum, and the associated size and magnetic field in this minimum energy (and close to equipartition) case. Deviations from this minimum energy (equipartition) scenario are of course possible within the homogeneous spherical model. This means the true values of the size and magnetic field in the emitting region may well differ from those given by equations \eqref{eq:r_homog_nu_form} and \eqref{eq:B_homog_nu_form} (see Appendix~\ref{sec:equipartition_deviation}). However, the dependence of the minimum energy on size is extremely strong as seen in equation \eqref{eq:enrgy_dependence_size}. This means in reality, as often external energetic constraints are available, the size of the emitting region can be very accurately calculated (assuming this stationary homogeneous quasi-spherical model). A change in size from the equipartition value of less than an order of magnitude results in a required energy which likely violates external constraints. However, the magnetic field is much less strongly constrained by the equipartition estimate, due to the weaker dependence of the energy on $B$.

We also note that even in the case where the SSA peak is not observed and instead optically thin emission is seen equations \eqref{eq:energy_homog_nu_form} to \eqref{eq:B_homog_nu_form} can still give useful constraints on the physical conditions in the emitting region. If optically thin emission of flux density $F_{\nu_\text{obs}}$ is seen at an observing frequency $\nu_\text{obs}$, then from the shape of the synchrotron spectrum we know $\nu_1 < \nu_\text{obs}$ and $F_{\nu_1} > F_{\nu_\text{obs}}$. From the dependencies of $F_{\nu_1}$ and $\nu_1$ in \eqref{eq:energy_homog_nu_form} to \eqref{eq:B_homog_nu_form}, we see that the above constraints enable us to use $F_{\nu_\text{obs}}$ and $\nu_\text{obs}$ in these formulae to obtain lower bounds on the minimum energy and associated size, and an upper bound on the associated magnetic field. If the energy obtained by using $F_{\nu_\text{obs}}$ and $\nu_\text{obs}$ in \eqref{eq:energy_homog_nu_form} is $E_\text{obs}$, then this is related to the true $E_\text{homog,min}$ (in the frequency cut-off formalism) by:


\begin{equation}
    E_\text{homog, min} = E_\text{obs} \left(\frac{\nu_\text{obs}}{\nu_1}\right)^{\frac{45-18\alpha}{34}} , 
\end{equation}

\noindent demonstrating that $E_\text{obs}$ is an underestimate of the true energy, with the exact magnitude of the underestimate depending on how far $\nu_\text{obs}$ is from $\nu_1$ and the value of $\alpha$ (or equivalently $p$). Note that in deriving the above we have assumed $p=1-2\alpha$. A similar conclusion is reached for the associated size where $R_\text{obs}$ is the size obtained by using $F_{\nu_\text{obs}}$ and $\nu_\text{obs}$ in \eqref{eq:r_homog_nu_form}:

\begin{equation}
    R_\text{homog, eq} = R_\text{obs} \left(\frac{\nu_\text{obs}}{\nu_1}\right)^{\frac{35-14\alpha}{34}} . 
\end{equation}

\noindent A use case for these formulae are flares from unresolved radio transients which are observed to remain optically thin throughout. As an example these events often occur in X-ray binaries (e.g. \citealt{fender_bright_2019}, \citealt{fender_2023}). As the emitting regions are not resolved, equipartition estimates using the volume are not applicable. Instead one of the few ways of constraining the size and energy of the emitting region is through the formulae above.

The formulae derived in this Section assume a stationary emitting region, e.g. one with no bulk relativistic motion (or significant cosmological redshift). Due to Doppler boosting, bulk relativistic motion affects the observed flux density and the observed peak frequency of the synchrotron spectrum. It is then necessary to correct SSA estimates if the bulk relativistic motion and angle of the motion to the line of sight is known. These corrections are outlined in Appendix~\ref{sec:doppler_and_redshift}. $E_\text{homog,min}$ and $R_\text{homog,eq}$ derived using the equations in this Section can also be taken as $E_\text{eq,N}$ and $R_\text{eq,N}$ and corrected for bulk relativistic motion in the specific ultra-relativistic blast wave model of \cite{bnp_2013} and \cite{matsumoto_2023}, bearing in mind caveats discussed in Section~\ref{sec:comparison_to_prev}. The derived $E_\text{homog,min}$ and $R_\text{homog,eq}$ combined with a timescale can also be used to constrain angle to the line of sight and bulk relativistic motion in both a  model dependent and independent manner as demonstrated by e.g. \cite{bright_2025}.


\subsection{Comparison to previous formulae}
\label{sec:comparison_to_prev}
Equations \eqref{eq:energy_homog_nu_form} to \eqref{eq:B_homog_nu_form} are similar to equations~24 to 26 from \cite{fender_bright_2019}, but without explicit dependence on the integrated synchrotron luminosity, $L$, and with full dependence on $p$. Both $L$ and $c_{12}$ in the equations from \cite{fender_bright_2019} have implicit dependencies on several observed factors, including the details of the electron energy distribution.

We simulate synchrotron spectra with known parameters (assuming close to equipartition), and confirm that our equations \eqref{eq:energy_homog_nu_form} to \eqref{eq:B_homog_nu_form} (combined with \ref{numax_and_nu1} and \ref{eq:fnu1_and_fpeak}) recover the correct physical parameters. Due to only including partial $p$ dependence (in $c_{12}$ and $L$), equations~24 to 26 from \cite{fender_bright_2019} can have errors of up to a factor of a few for extreme values of $p$ over the range from 1.5 to 4. We note that values of $p$ at these extremes have been inferred from some observations (e.g. \citealt{hughes_2025}, \citealt{tetarenko_2017}) and previously \cite{bjornsson_2021} noted that for some methods assuming $p=2$ when this is not the case can lead to order of magnitude errors.

However, the scenario changes when $c_{12}$ and $L$ are not calculated accurately, and instead are approximated e.g. the commonly used $L=4 \pi D^2 F_{\nu} \nu$. In these case, independent of $p$, errors of over three orders of magnitude are possible when compared to our formulae, and the true values used to generate synchrotron spectra. Errors are particularly large when $c_{12}$ and $L$ are computed using inconsistent values of $\nu_\text{min}$, $\nu_\text{max}$ or $p$. This is the case for the single frequency approximations (equations 28-31) in \cite{fender_bright_2019} (as noted by the authors themselves). We therefore caution against making these approximations, and instead argue that even if $\nu_\text{min}$ and $\nu_\text{max}$ (or equivalently $E_\text{min}$ and $E_\text{max}$ in the $K_E$ formalism) are unconstrained by observations, this uncertainty should be taken into account fully while making estimates using the SSA peak. Bounds on these quantities can almost always be found through physical arguments (e.g. see equation \ref{eqn:gamma_min_constraint}).

We can also compare our formulae to similar equations in \cite{bnp_2013} and \cite{matsumoto_2023}. There are two key differences in this case. 

First, the formulae in \cite{bnp_2013} and \cite{matsumoto_2023} necessarily assume a specific, geometric model relating the radius and depth to the Lorentz factor, originating from the ultra-relativistic blast wave model presented in \cite{blandford_mckee_1976}. Furthermore, the model in \cite{bnp_2013} and \cite{matsumoto_2023} necessarily makes the assumption of a freely expansing outflow, so that the distance the synchrotron emitting region has travelled from some origin and the size of the emitting region are linearly related. These assumptions are required so that relativistic sources where the the inclination and bulk Lorentz factor cannot be determined independently can be constrained from the lightcurves. However in some scenarios, for example where outflows can be resolved (e.g. \citealt{miller_jones_2019}), the inclination and bulk Lorentz factor can be independently measured and so the relativistic blast wave model assumptions are not strictly needed. Instead the SSA formalism presented in this work could be used (including, where appropriate, the relativistic corrections presented in Appendix~\ref{sec:doppler_and_redshift}), to obtain a more model independent measure of the physical quantities probed by SSA. We note that applying the formalism in this work and in \cite{matsumoto_2023} together in cases where independent constraints are available, would also allow for the assumptions of the ultra-relativistic blast wave model to be tested (so long as differences in the treatment of the electron energy estimates are accounted for; see the next paragraph for discussion of this). Overall, the scaling of the formulae in \cite{bnp_2013} and \cite{matsumoto_2023} with the observed quantities in the non-relativistic limit are reproduced by the formulae in this work for the $p=2$ case. In the relativistic case (Appendix~\ref{sec:doppler_and_redshift}), the scalings are somewhat different from those in \cite{bnp_2013} and \cite{matsumoto_2023} due to the aforementioned model assumed in these works.

Secondly, both \cite{bnp_2013} and \cite{matsumoto_2023} only take into account energy contributions from electrons radiating at $\nu_\text{min}$. Despite the fact that these electrons radiating at $\nu_\text{min}$ do dominate the total energy contribution, neglecting to account for electrons throughout the power law up to the high energy cut-off can lead to underestimation of the minimum energy estimate by over an order of magnitude for modest ranges of $\frac{\gamma_\text{max}}{\gamma_\text{min}}\sim10^2$ for $p\sim2$. Electron energy spectra with cut-offs spanning a larger range will result in greater underestimates. This assumption also leads to underestimation of magnetic field strengths by a factor of a few but only has an effect of order unity on the emitting region sizes due to the weak dependence on $K_E$ or $K_{\nu}$ as seen in e.g. equation \eqref{eq:r_homog_nu_form}.



\subsection{Correcting for deviations from sphericity}\label{sec:results_spehrical}

The formulae outlined in Section~\ref{sec:trad_estimates} to obtain the minimum energy and associated physical parameters are only valid under the assumption of quasi-sphericity of the emitting region. For a general emitting region geometry, quasi-sphericity is the case where the observed solid angle \textit{and} volume of the emitting region can be written as $\frac{\pi R^2}{D^2}$ and $\frac{4}{3} \pi R^3$, where in this case $R$ is some scale size. As previously noted, for a cylindrical slab this condition is satisfied when the depth of the slab is related to the radius as $s=\frac{4}{3}R$. 

We can consider the impact of deviating from the assumption of quasi-sphericity on traditional SSA energy estimates. We do this by analytically exploring face-on cylindrical slab models where quasi-sphericity is no longer assumed. Relaxing the assumption of quasi-sphericity leads to degeneracy, as the radius and depth of the emitting region are no longer linked as $s=\frac{4}{3}R$. The result is that an identical synchrotron spectrum, such as the one shown in Figure~\ref{fig:default_sync_spec}, can be generated both by a cylinder with a large depth compared to radius (a rod-like geometry), and a lower magnetic field, or a cylinder with a low depth compared to radius (a pancake-like geometry) and a higher magnetic field. In fact, the same synchrotron spectrum can be reproduced by a cylindrical slab with \textit{any} desired ratio of depth to radius $q=\frac{s}{R}$, so long as appropriate adjustments in magnetic field strength are made. The physical parameters of the emitting region which produces the observed spectrum, such as the size, magnetic field, and energy, depend on $q$. The results for $q=\frac{4}{3}$ are what is given by traditional SSA estimates of physical parameters.

By requiring that $\nu_1$ and $F_{\nu_1}$ remain constant, assuming equipartition ($E_e \sim E_B$), and varying only $B$, $s$, and $R$, we can derive the dependence of the energy on $q$, for the set of models which produce the same observable spectrum. Equipartition in the energy cut-off formalism\footnote{In the frequency cut-off formalism the deviation from quasi-sphericity correction factors derived in this Section will be slightly different, as equipartition instead leads to $N_0 \propto B^{\frac{6-p}{2}}$.} (i.e. holding the electron energy cut-offs constant) gives us that $N_0 \propto B^2$. Then from equation \eqref{fnu1_eqn}, requiring $\nu_1$ and $F_{\nu_1}$ are constant, we obtain $R \propto B^\frac{1}{4}$. From equation \eqref{nu1_eqn} we obtain that $s \propto B^\frac{-(p+6)}{2}$. Combining these we obtain that in order for the observed spectrum to remain unchanged while varying the deviation from sphericity, $q \propto B^{\frac{-2(p+6)-1}{4}}$ must hold. Then from equation \eqref{mag_energy} for the magnetic energy of a homogeneous emitting region, we obtain $E \propto s R^2 B^2$. Eliminating $B$ we obtain the relationship between the energy of the emitting region and the deviation from sphericity:

\begin{equation}\label{E_with_sphericity}
    E \propto q^\frac{2(1+p)}{2(p+6)+1} .
\end{equation}

\noindent For $p=2$ this gives $E \propto q^\frac{6}{17}$. This means that applying traditional SSA methods, which assume a quasi-spherical emitting region, will underestimate the energy in the case of a rod-like  geometry ($q>1$), and overestimate the energy in the case of a pancake-like geometry ($q<1$). Other physical parameters depend on the geometry in different ways:

\begin{equation}
    B \propto q^{-\frac{4}{2(p+6)+1}} ,
\end{equation}

\begin{equation}
    R \propto q^{-\frac{1}{2(p+6)+1}} ,
\end{equation}

\begin{equation}
    s \propto q^{\frac{2(p+6)}{2(p+6)+1}} ,
\end{equation}

\begin{equation}\label{V_with_sphericity}
    V \propto q^{\frac{2(p+5)}{2(p+6)+1}} .
\end{equation}

\noindent In many cases one cannot be certain of the emitting region geometry, and when it is possible to resolve emitting regions deviations from spherical geometries have been observed (e.g. \citealt{bahramian_2023}). Therefore, care should be taken when applying SSA methods to estimate physical parameters to fully appreciate the uncertainty in the emitting region geometry, and the relevant uncertainty on the derived parameters. For example, when using traditional SSA techniques the calculated $R$ and $E$ are somewhat weakly dependent on the source geometry deviating from sphericity, whereas, $s$ and $V$ have significantly stronger correction factors (see equations~\eqref{E_with_sphericity} to \eqref{V_with_sphericity}). If \textit{a priori} information about the geometry of the emitting region is known, then correction factors to traditional SSA methods can be calculated using the equations above, by taking the ratio of the known $q$ and $q=\frac{4}{3}$ to the relevant power. We note that the dependencies of physical parameters on $q$ above in the case $p=2$, agree with the equivalent dependencies on geometry in the Newtonian case in \cite{bnp_2013}.

Physical scenarios for both rod-like and pancake-like geometries could be expected in astrophysical sources, and in fact the same scenario can give rise to both of these geometries, depending on the observer viewing angle. For example, a shock front created by a blast wave where there is a thin layer of synchrotron emitting plasma may have a pancake-like geometry at low inclination angles (measured relative to the bulk velocity of the material) but a rod-like geometry (meaning a large depth along the line of sight when compared to area on the sky) when viewed at high inclination angles. Therefore, for a given population of sources, or even the same source if for example precession is known to occur (e.g. \citealt{cowie_2025}) then the effects of deviations from sphericity could introduce spurious inclination dependent effects on the measured parameters from SSA. These effects could compete or compound with related effects from relativistic bulk motion due to the dependence of the Doppler factor on inclination to the line of sight, and an investigation into the effect of this on SSA estimates is left for future work. A good example of such a population are XRBs, which show SSA flares due to the launch of discrete jets \citep{fender_bright_2019}, but as a sample are isotropically selected based on their X-ray emission.

\section{Motivation for inhomogeneity}\label{sec:motivation}

There were several approximations made when outlining the model and energetic estimates in Section~\ref{sec:homog}, which are not thought to hold in the astrophysical environments this model is applied to. Several of these approximations would effect the energetic estimates from SSA methods, for example the presence of baryons or thermal electrons (Appendix \ref{sec:equipartition_deviation} and e.g. \citealt{bnp_2013}), or relativistic bulk motion of the plasma (Appendix \ref{sec:doppler_and_redshift} and e.g. \citealt{fender_bright_2019}). Here we focus on discussing the relaxation of those specific approximations which can lead the flattening of the spectral index below the spectral peak, often observed in synchrotron sources, as outlined in Section~\ref{sec:intro}. This behaviour is not possible within the simple model discussed in Section~\ref{sec:homog}.

\subsection{Electron energy distribution effects}

Firstly, as noted in Section~\ref{sec:homog}, the non-thermal distribution of ultra-relativistic electrons will not be unbound. Both the high and low energy cut-offs have distinct effects on the idealised spectrum in Figure~\ref{fig:default_sync_spec}. The high energy cut-off alters the spectrum such that above the frequency given by $\nu_{cr}(E_{\text{max}},B,\theta)$ there will be an exponential cut-off in the synchrotron spectrum. The effect of the low energy cut-off is more nuanced and depends on the relationship between $\nu_\text{min} = \nu_{cr}(E_{\text{min}},B,\theta)$ and $\nu_1$ (the frequency where the optical depth of the emitting region is unity, close to the SSA peak) as we discuss below and as summarised in e.g. \cite{granot_van_der_horst_2014}.


For the case where $\nu_\text{min} < \nu_1$ the spectrum is unchanged around the SSA peak. However, below the frequency $\nu_\text{min}$ the emission is dominated by emission from the lowest energy electrons. These are at a single energy and so the self absorbed spectral index becomes $+2$ (see equation 3.48 of \citealt{pacholczyk}). 

For the case of $\nu_\text{min} > \nu_1$ we observe at frequencies below $\nu_1$ a self absorbed spectrum with an index of $+2$ as we are again dominated by the lowest energy electrons. Between $\nu_1$ and  $\nu_\text{min}$ we see a power law with slope $+\frac{1}{3}$, as we are still dominated by lowest energy electrons but absorption is not important. Then above $\nu_\text{min}$ we have an optically thin spectrum from a power law distribution of electrons with a spectral index of $\frac{1-p}{2}$. Therefore, for the case $\nu_\text{min} > \nu_1$ spectral indices below the peak in the spectrum (now the peak is not caused by SSA) are flatter than those when the peak is due to SSA. In this case the formulae in this work are not applicable. For example the emission and absorption coefficients (equations \ref{emmission_eq} and \ref{absorption_eq}) are for a power law population of electrons instead of a mono energetic population. The equivalent equations for this case can be found in \citep{pacholczyk}. 

However, the case $\nu_\text{min} > \nu_1$ cannot explain spectral indices between $+\frac{1}{3}$ and $+2$ below the spectral peak, which are most often observed, without invoking additional synchrotron emitting components. In fact, if the peak in the spectrum can be identified as due to SSA (e.g. through a spectral index $>\frac{1}{3}$), then this can be used to constrain the minimum energy cut-off in the electron energy spectrum (see equation~\eqref{eqn:gamma_min_constraint}). Throughout this work, we will assume that the cut-off energies $E_\text{max}$ and $E_\text{min}$ correspond to frequencies outside of the frequency range observed, and we will additionally neglect synchrotron cooling, which typically only effects the optically thin part of the spectrum \citep{sari_1998}.

Flattening of the spectral index below the SSA peak can also be achieved if instead of following a non-thermal power law, the electron energy distribution is a thermal relativistic Maxwellian distribution. In this case, the spectral index below the spectral peak is $+2$, and this thermal synchrotron peak may be significantly broadened compared to a typical SSA peak, if the SSA frequency is close to the typical synchrotron frequency of the electrons in the Maxwellian distribution (\citealt{jones_1979}, \citealt{margalit_2021}). This broadened peak may manifest as a measured flattened spectral index below the spectral peak. However, a thermal relativistic population of electrons are expected to produce a steep negative spectral index ($\alpha<-1.5$) at frequencies above the SSA peak for a wide range of parameters \citep{margalit_2021}, and this is not often observed.

A flattened spectral index can also be obtained by a purely optically thin component where $p < 1$. However, this is unlikely given our understanding of particle acceleration processes (\citealt{fender_2001}, \citealt{matthews_2020}).




\subsection{Observational effects}

Observational effects which are not taken into account in the simple model can also lead to flattening of the spectral index below the spectral peak. It is worth summarising the 3 main types of spectral index measurement made. A two point spectral index is the most common measurement made, due to its convenience. It is made by taking the flux measurements, typically from two separate instruments, at 2 frequencies, and assuming a power law connects the two points. The weakness of 2 point spectral indices is that features in the spectrum which deviate from a power law cannot be measured, and will likely bias the measurement. Multi-band spectral indices typically use multiple flux measurements from multiple facilities, and a power law fit. Finally, in-band spectral indices are measured by breaking up spectral information from a single instrument and fitting a power law. These are similar to multi-band spectral indices in their advantages, but they differ as they tend to cover a smaller range of frequencies, but have higher spectral resolution (dependent on the instrument). In-band spectral indices can exhibit biases towards flatness when frequency integrated signal to noise is below $\sim35$, but at high signal to noise ratios are accurate (e.g. \citealt{heywood_2016}).

A flatter spectral index will be measured if one or more points used to calculate the spectral index lie within the frequency range close to the SSA peak. To understand whether this can have a significant effect we define the frequency range of the SSA peak as the part of the spectrum which deviates from the power law approximations by $>5\%$. This part of the spectrum covers an range in frequency of a factor $\sim3$ and this value is only weakly dependent on $p$. Therefore, in order for the SSA peak to significantly bias a spectral index measurement of emission the SSA peak must be within a factor of $\sim1.5$ in frequency from the highest frequency used to calculate the spectral index. Furthermore, in-band spectral indices and multi-band spectral indices are less susceptible to this bias. If the fractional width in frequency covered by the measurements is large enough, deviation from a power law can be seen in the spectrum and accounted for, in order to measure the true spectrum below the SSA peak.


\begin{figure}
 \includegraphics[width=\columnwidth]{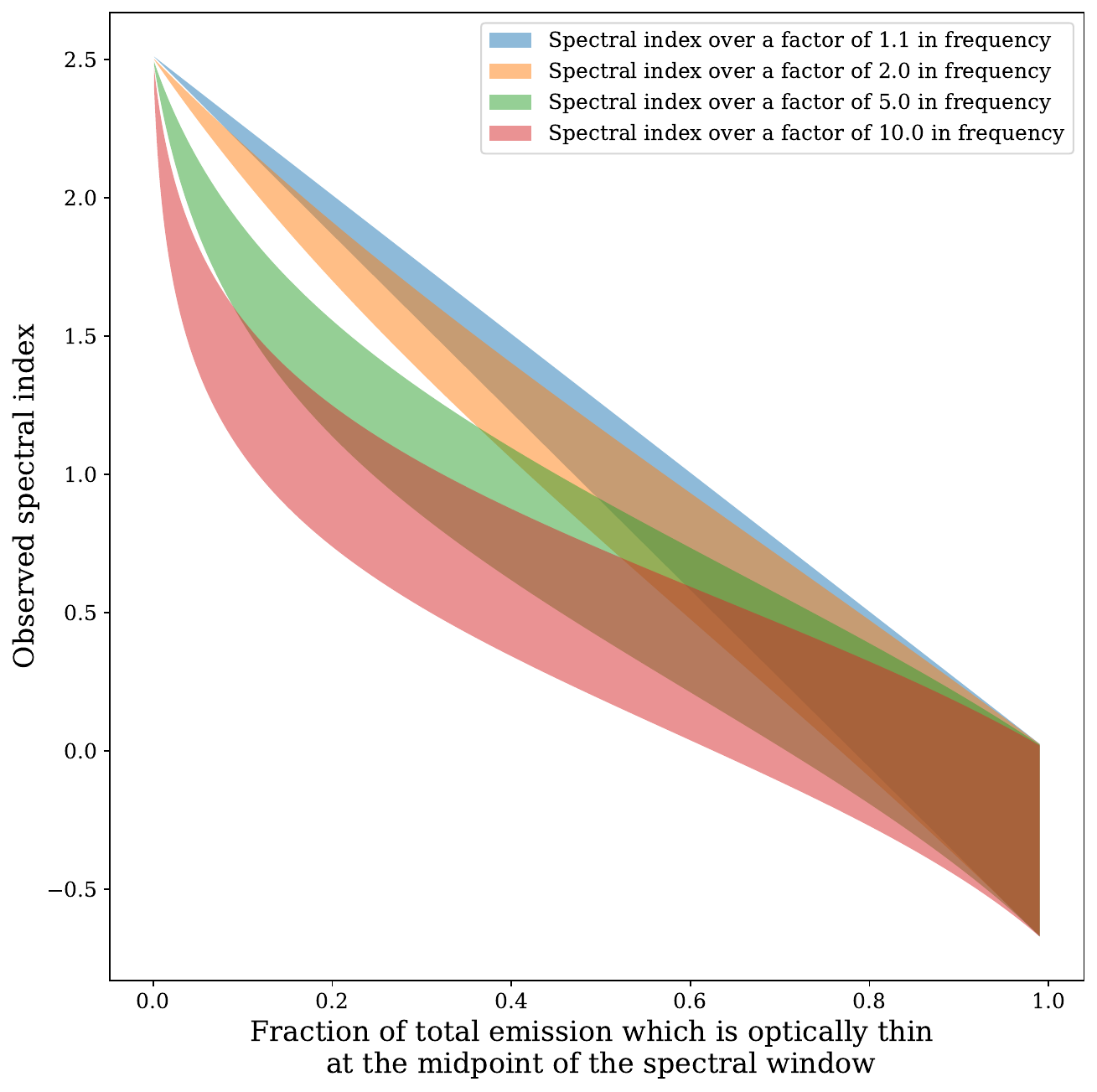}
 \caption{The two-point spectral index of an optically thick synchrotron source with contaminating optically thin emission which makes up some fraction of the total emission. The different coloured curves represent measuring the spectral index over different ranges in frequency. The spread in the curves represents the range of possible spectral indices for the optically thin component, from $0$ to $-0.7$. The fraction of total emission from the optically thin component is measured at the midpoint of the spectral window considered.}
 \label{fig:frac_optically_thin}
\end{figure}

The second observational effect, is that due to spatial resolution limitations, the measured flux may not be dominated by a single isolated synchrotron source. For example, if there is one or more \textit{independent} optically thin synchrotron sources contributing to the spectrum alongside an optically thick source, the spectral index below the SSA peak will be altered. In order to investigate the importance of this effect we simulated a single optically thick and a single optically thin component of different fluxes. The optically thin component was allowed to have a spectral index ranging between $0$ and $-0.7$. We then measured the spectral index using a two point spectral index over frequency ranges of different sizes. Figure~\ref{fig:frac_optically_thin} shows the results of this simulation with the observed 2 point spectral index over the different frequency ranges, on the y-axis, where the spread corresponds to the optically thin components with different spectral indices. On the x-axis is the fraction of the total emission which is optically thin, based on fluxes at the midpoint of the spectral window. From this we can see that to produce a substantial flattening of the measured two-point spectral index, e.g. to $\sim 1$, either the two-point spectral index must be taken over a very large range in frequency, or, the optically thin emission component(s) must dominate the flux. Once again, in-band and multi-band spectral indices, providing a large enough fractional width in frequency covered, are less susceptible to this effect. Noticeable curvature will be present in these measured spectra as at low frequencies one would expect to see the spectrum turn over and the spectral index to become negative as the optically thin component dominates. In many scenarios the presence of a dominant optically thin component can be ruled out, either by never having observed a low frequency turnover to optically thin emission, by variability arguments, or using physical arguments, as the presence of a dominant optically thin components requires a large, energetic, emitting region.


We argue that in many cases, the most likely explanation for a spectral index below the peak in the spectrum which is inconsistent with the $+2.5$ expected from a simple model, is inhomogeneity. Based on observations of a synchrotron spectrum one can decide if any of the other arguments above apply, and in fact the two emission component case discussed above could be seen as a special case of inhomogeneity where the components are independent of each other. Therefore, in the next Section, we use some general models of inhomogeneity to explore the impact on the observed spectrum, and on the physical parameters measured using SSA methods.

\section{Model framework}\label{sec:model}

We use two simple models with different geometries to investigate the impact of inhomogeneity on energy estimates from SSA observations. The first is these is a power-law inhomogeneous cylindrical slab. This model was first used to investigate the effects of inhomogeneity on the synchrotron spectrum by \cite{de_bruyn_1976}. The second of these is a broken power-law inhomogeneous sphere, implemented using the \textsc{flaremodel} package \citep{dallilar_2022}. We work in the energy cut-off formalism for all models, i.e. holding the electron energy cut-offs constant within and between models.




\begin{figure}
 \includegraphics[width=\columnwidth]{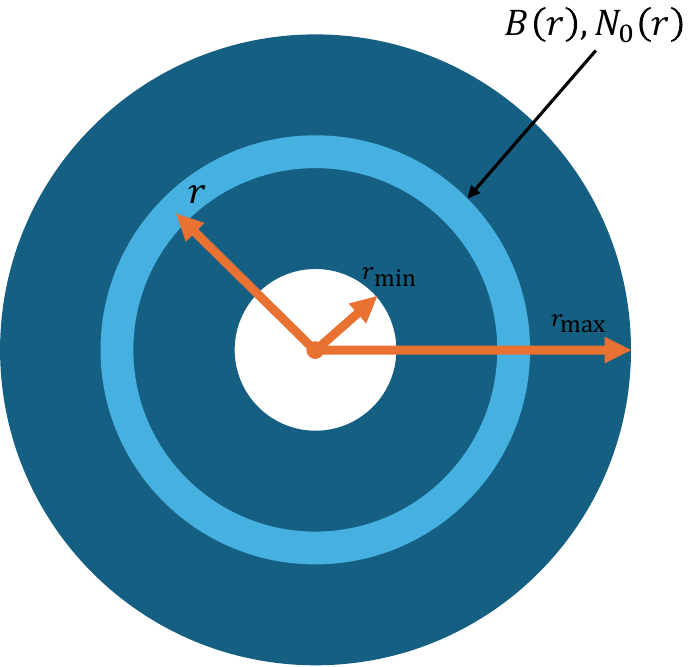}
 \caption{The view of the observer (face on) of the inhomogeneous cylindrical shell slab model used extensively throughout this work. The cylindrical shell is parametrised by an inner and outer radius, $r_{\text{min}}$ and $r_{\text{max}}$ respectively, and a depth (into the page), $s$. The geometry is split into a series of cylindrical shells of infinitesimal radial thickness through which the magnetic field and synchrotron emitting particle density are constant. This means there is no inhomogeneity along the line of sight (into the page). The details of how the magnetic field and synchrotron emitting particle density vary with the radial coordinate are presented in the text.}
 \label{fig:model_diagram}
\end{figure}

\subsection{Inhomogeneous slab model}\label{sec:model_slab}

We assume a power law distribution in energy of synchrotron emitting electrons with an isotropic distribution of pitch angles. As previously stated we neglect the effects of cut-offs in the electron energy spectrum, and we neglect the effect of synchrotron cooling. We use the fiducial value of $p=2$ for the electron energy index.

The model has a cylindrical shell geometry, where the observer views one of the faces of the cylindrical shell, with outer radius $r_{\text{max}}$, and inner radius $r_{\text{min}}$. The depth of the cylinder is given by $s$, which we take to be $\frac{4}{3} (r_{\text{max}}^2 - r_{\text{min}}^2)^\frac{1}{2}$, so that the cylindrical shell is always quasi-spherical. We then allow both the magnetic field and $N_0$ parameter to vary axisymmetrically, as a function of the coordinate $r$, while being independent of depth along the cylindrical shell. Therefore, the shell is homogeneous along the line of sight of the observer, while being inhomogeneous in the plane perpendicular to this. This model geometry is shown in Figure~\ref{fig:model_diagram}. The coordinate $r$ represents a position within the emitting region, rather than the distance from some origin for an expanding blast wave of similar, as is often used in some literature. 


We parametrise the magnetic field and $N_0$ as power laws:

\begin{equation}\label{B_with_r_eq}
    B(r) = B(r_{\text{min}}) \left(\frac{r}{r_{\text{min}}}\right)^{-m} ,
\end{equation}

\begin{equation}\label{N_with_r_eq}
    N_0(r) = N_0(r_{\text{min}}) \left(\frac{r}{r_{\text{min}}}\right)^{-n} .
\end{equation}


\noindent The inner radius, $r_{\text{min}}$ is present to avoid singularities, and in reality represents the approximate size scale for an inner homogeneous core. In the case we will focus on where $m$ and $n$ are positive, any addition of a homogeneous core region will have a limited impact on the spectrum (\citealt{condon_1973}, \citealt{marscher_1977}). We verify this for cylindrical slab model by repeating our analysis outlined later in this Section with the addition of a homogeneous core. We find that this effects our correction factors discussed in Section~\ref{sec:results} always by less than $\sim50\%$ over the whole parameter range, and by much less over most of the parameter range. In particular, effects are only seen for inhomogeneous models with large observed spectral indices and low observed translucent frequency decades (e.g. the bottom right corner of Figure~\ref{fig:energy_ratios_contour}). We also find that the addition of a homogeneous core can both increase and decrease  correction factors depending on the exact model parameters.

This model setup allows us to treat this inhomogeneous cylindrical shell as a series of infinitesimally thin (in the $r$ direction) homogeneous cylindrical shells, each with a magnetic field and $N_0$ given by the above equations, and with the constant depth. To allow for fair comparison between the inhomogeneous model and homogeneous model we ensure that the ratio of magnetic and particle energy at $r_\text{min}$ is the same as the ratio in the homogeneous model, $\frac{E_B(r_\text{min})}{E_e(r_\text{min})} = \frac{E_{B,\text{homog}}}{E_{e,\text{homog}}}$. We take this ratio to be $\frac{6}{11}$, which is the ratio close to equipartition which gives the minimum energy for the unknown size case (see Section~\ref{sec:trad_estimates}). However, varying this ratio, so long as it is the same between the compared homogeneous and inhomogeneous model has no effect on any of our results.

If $n$ and $m$ are independent parameters there is no guarantee that this ratio between magnetic and particle energy will remain constant throughout the inhomogeneous source. The requirement to maintain this energy ratio, assuming that the electron spectrum energy cut-offs are kept constant over the inhomogeneous source (i.e. the energy cut-off formalism), is that $n=2m$. We find that any deviation from this raises the energy of the inhomogeneous model considered, and hence raises the energy correction factors discussed in Section~\ref{sec:results}. The reason for this is that any deviation from $n=2m$ will inevitably lead to parts of the inhomogeneous source which are far from equipartition, and therefore require a larger energy to produce a given observed synchrotron spectrum. This leads to the conclusion that the energy correction factors presented in Section~\ref{sec:results} are best thought of as the \textit{minimum} correction factors to the traditional (homogeneous) SSA minimum energy estimate for an inhomogeneous source. 


The spectrum of the inhomogeneous cylindrical shell, $F_{\nu,\text{inhom}}$ can then be calculated by integrating the spectra of the many infinitesimal cylindrical shells:

\begin{equation}\label{inhomog_integral}
    F_{\nu,\text{inhom}} = \int^{r_{\text{max}}}_{r_{\text{min}}} 2 \pi r \frac{1}{D^2} I_\nu(r) dr ,
\end{equation}

\noindent where $I_\nu$ is given by equation \eqref{intensity_eq} and depends on $r$ for the inhomogeneous model. The energy of the inhomogeneous cylindrical shell can be calculated in a similar manner, but also has an analytical solution when $m \neq 1$:

\begin{align}\label{eqn:inhomog_energy}
    E_{\text{inhom}} = & \frac{17}{11} \frac{4}{3} 2 \pi K_E N_0 (r_\text{min}) \nonumber \\ &\frac{r_\text{min}^{2m} \left(r_\text{max}^2 - r_\text{min}^2\right)^{\frac{1}{2}}}{2-2m} \left(r_\text{max}^{(2-2m)} - r_\text{min}^{(2-2m)}\right) .
\end{align}

\noindent To generate the spectrum from a inhomogeneous synchrotron slab we compute the integral in equation \eqref{inhomog_integral} using the {\sc{scipy.inegrate.quad}} method \citep{scipy} which performs adaptive integration using Gauss-Kronrod rules \citep{quadpack}. For cases where $m>1$, the integrand is sharply peaked in $r$ space and so we change the upper limit of the integral to be 1 order of magnitude larger than the location of the peak in the integrand in order to improve computation time. To evaluate the whole spectrum we compute the integral at $10^5$ frequency values logarithmically spaced between upper and lower limits. Changes in the effective upper limit of the integral as described above, and in frequency resolution, as well as other integration settings had no effect on our results when investigated.

To understand what effect inhomogeneity has when traditional SSA estimates are applied to an inhomogeneous source we then compare the energies of homogeneous and inhomogeneous cylindrical slab models which have the same key observable parameters. For a given inhomogeneous cylindrical slab spectrum generated as described above, we use the traditional SSA formulae in the energy cut-off formalism (equations \ref{eq:e_homog_E} to \ref{eq:r_homog_E}) to find the energy and associated size and magnetic field if one (incorrectly) assumes homogeneity\footnote{The requirement to have exactly accurate traditional SSA estimations so that the homogeneous counterpart to the inhomogeneous spectrum could be found is what initially motivated the derivation outlined in Section~\ref{sec:trad_estimates}.}. We then compare these to the energy, size and magnetic field used to generate the inhomogeneous spectrum. The ratio of $E_\text{inhomog}$ and $E_\text{homog}$ is then the energy correction factor which must be applied to traditional SSA estimates in order to account for inhomogeneity. Similarly the ratio of $(r_{\text{max}}^2 - r_\text{min}^2)^\frac{1}{2}$ and $R_\text{homog}$ is the size correction factor. We note that $(r_{\text{max}}^2 - r_\text{min}^2)^\frac{1}{2} \rightarrow r_\text{max}$ as $r_\text{max} >> r_\text{min}$, which is the case for most of the parameter space explored. We find that these energy and size correction factors depend only on the choice of $\frac{r_{\text{max}}}{r_{\text{min}}}$ and $m$ (as $n$ is constrained to be $n=2m$). Finally, simple checks show that as $m \rightarrow 0$, the correction factors $\rightarrow 1$ as expected.

\subsection{Inhomogeneous sphere model}\label{sec:model_spherical}

The power law inhomogeneous slab model is one of the simplest prescriptions of inhomogeneity, and while a powerful tool for understanding general effects, it is perhaps unlikely to arise in real astrophysical environments. Therefore, we chose to explore a potentially more physical model, which also allows us to investigate the effect of a different geometry. We model a spherical emitting region of size $r_\text{max}$, where $N_0$ and $B$ are a broken power law function of the radius $r$:

\begin{equation}
    B(r)=B(0) \left(1+\left(\frac{r}{r_\text{min}}\right)^w\right)^{-\frac{m}{w}} ,
\end{equation}

\begin{equation}
    N_0(r)=N_0(0) \left(1+\left(\frac{r}{r_\text{min}}\right)^w\right)^{-\frac{n}{w}} ,
\end{equation}

\noindent where $w$ is a sharpness parameter, which controls the sharpness of the transition between the constant ($r<r_\text{min}$) and power-law ($r>r_\text{min}$) regimes. We take a value of $w=10$ to give a sharp cut-off for reasons discussed in Section~\ref{sec:results_slab_spec}. For the same reasons as in the slab model (discussed in Section~\ref{sec:model_slab}) we take the ratio of magnetic to particle energy at $r=0$ to be $\frac{6}{11}$ and we enforce $n=2m$, meaning the calculated energy correction factors can be considered as a minimum.

This spherical model then represents a power-law sphere with a homogeneous core with size $r_\text{min}$. To calculate the energy of the emitting region in this model we numerically integrate:

\begin{equation}
    E_\text{inhomog, sph} = \frac{17}{11} 4\pi K_E N_0(0) \int^{r_\text{max}}_0 r^2 \left(1+\left(\frac{r}{r_\text{min}}\right)^w\right)^{-\frac{n}{w}} dr ,
\end{equation}

\noindent using the \textsc{scipy.integrate.quad} method \citep{scipy}. To calculate the spectrum from this spherical model we use the radial sphere implementation of the \textsc{flaremodel} package outlined in \cite{dallilar_2022}. Specifically we generate the spectrum over 300 frequency values using 400 radial grid points.

With the energy of the inhomogeneous sphere, we then follow the same process for calculating the energy and size correction factors as described in Section~\ref{sec:model_slab}. Once again we find these correction factors only depend on the choice of $\frac{r_{\text{max}}}{r_{\text{min}}}$ and $m$, due in part to our choice of large $w$ which is discussed in the next Section. 

\section{Results}\label{sec:results}

\subsection{Spectrum of an inhomogeneous source}\label{sec:results_slab_spec}

\begin{figure}
 \includegraphics[width=\columnwidth]{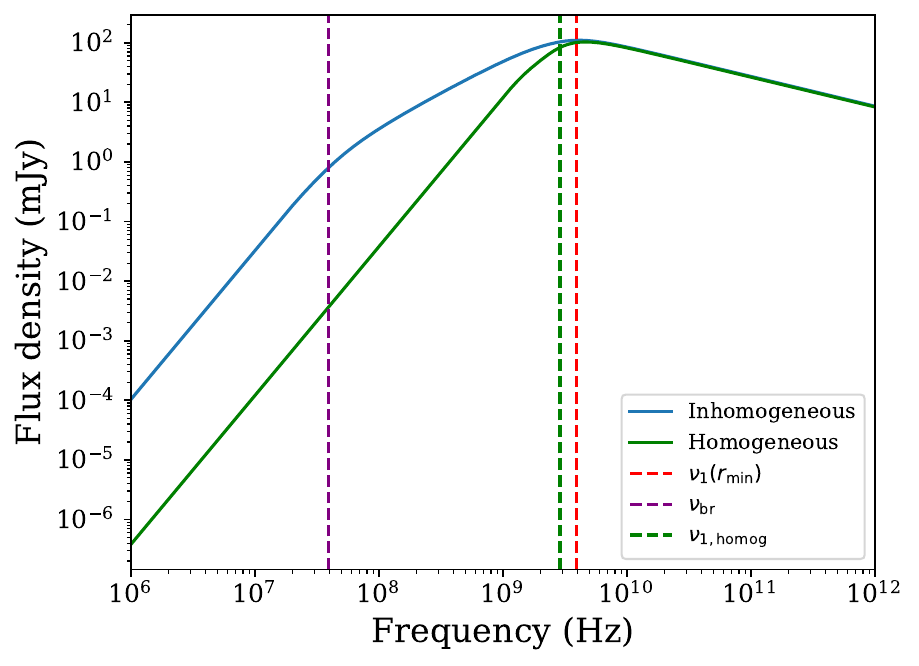}
 \caption{Fiducial inhomogeneous spectrum generated by the model described in Section~\ref{sec:model} and the equivalent homogeneous spectrum which reproduces the position and amplitude of the spectral peak. The model parameters used to generate the spectrum are $m=1.5$, $r_{\text{min}} = 5 \times 10^{12} \: \text{cm}$, $r_{\text{max}} = 5 \times 10^{13} \: \text{cm}$, $B(r_\text{min})=1\:\text{G}$ and a distance of 1 kpc. $\nu_1$ for both the equivalent homogeneous model and the for the emitting region at $r_\text{min}$ in the inhomogeneous emitting region are shown.}
 \label{fig:spectrum_comparison}
\end{figure}

We begin by analysing a fiducial spectrum generated by the inhomogeneous slab model in Section~\ref{sec:model_slab}, and comparing this with the equivalent homogeneous model which reproduces the position and amplitude of the spectral peak. Figure~\ref{fig:spectrum_comparison} shows the spectrum generated by an inhomogeneous cylindrical slab close to equipartition throughout where $m=1.5$, $r_{\text{min}} = 5 \times 10^{12} \:\text{cm}$, $r_{\text{max}} = 5 \times 10^{13} \:\text{cm}$, at 1~kpc, with a magnetic field at $r_{\text{min}}$ of 1~G. The electron energy distribution cut-offs are taken as an electron Lorentz factor of $10^{0.5}$ and $10^5$. We find the inhomogeneous spectrum has 3 distinct regions. 

The inhomogeneous spectrum peaks at a frequency close to $\nu_1$ evaluated at $r_{\text{min}}$, $\nu_1 (r_{\text{min}})$, which can be thought of as the SSA peak for the innermost regions of the source, and is hereafter referred to as the spectral peak. At frequencies much greater than $\nu_1 (r_{\text{min}})$, the whole source is optically thin and has the expected spectral index of $-0.5$ (for $p=2$). $\nu_\text{br} \equiv \nu_1 (r_{\text{max}})$ is the approximate frequency where a break from $+2.5$ to a flatter spectral index. For $\nu<<\nu_\text{br}$, the whole source is optically thick. For $\nu_\text{br} < \nu < \nu_1 (r_{\text{min}})$, inner parts of the source, with higher $B$-field and density of ultra-relativistic particles are optically thick while outer parts of the source remain optically thin. This part of the spectrum then has contributions from optically thin and thick emitting regions, creating a partially opaque or translucent part of the spectrum. This translucent part of the spectrum is at frequencies lower than the observed spectral peak. In the translucent part the spectral index is flatter than the optically thick spectral index of $+2.5$. Compared to the equivalent homogeneous spectrum, the inhomogeneous spectrum therefore shows significant excess emission below the observed spectral peak. 

The inhomogeneous spectrum therefore has two additional observable parameters when compared to the homogeneous model. The first of these is the spectral index in the translucent part, which for $p=2$ is given by the analytical formula (equation 7 in \citealt{de_bruyn_1976}):

\begin{equation} \label{transl_index}
    \alpha_\text{translucent} = \frac{3}{4} \left(\frac{17}{6} - \frac{2}{m}\right), 
\end{equation}

\noindent as long as $m>\frac{4}{7}$ \citep{de_bruyn_1976}. If $m$ is smaller than this then the above formula does not apply, and the spectrum typically no longer has 3 well defined regions, instead having a broader peak. However, here we are only interested in cases where $\alpha_\text{translucent} > 0$ and therefore $m>\frac{12}{17}$, as we typically observe spectral indices below the spectral peak $\gtrsim+0.4$. The second new observable parameter is the value of $\nu_\text{br}$. However, if the spectral peak is observed, then an equivalent observable is the range of frequency covered by the translucent part of the spectrum. As $\nu_\text{br} \equiv \nu_\text{br} \propto N_0 (r_{\text{max}})^\frac{2}{p+4} B(r_{\text{max}})^\frac{p+2}{p+4}$ from equation \eqref{nu1_eqn}, then we can show, using equations \eqref{B_with_r_eq} and \eqref{N_with_r_eq} that:

\begin{equation} \label{transl_freq_decades}
    \frac{\nu_1 (r_{\text{min}})}{\nu_\text{br}} = \left(\frac{r_{\text{max}}}{r_{\text{min}}}\right)^\frac{4m}{3} .
\end{equation}

\noindent Therefore, the range in frequency covered by the translucent part of the spectrum depends on both $m$ and $\frac{r_{\text{min}}}{r_{\text{max}}}$. Overall, the two new parameters introduced by the inhomogeneous model are completely constrained by the two new observables in the spectrum, and no degeneracy is present so long as quasi-sphericity and a constant relation between the magnetic and electron energy are assumed. 

We find that the spectrum of the broken power-law sphere described in Section~\ref{sec:model_spherical} has a virtually identical form to that of an inhomogeneous slab, with an optically thick, translucent and optically thin region. However, the spectral index in the translucent part has a different dependence on $m$ in the $p=2$ case. This translucent spectral index below the spectral peak and the frequency range over which it holds is not effected by the presence of the homogeneous core only for large values of the sharpness parameter $w$. This is because the transition region between the constant and power law regions of the emitting region is small for large $w$ leading to a minimal effect on the spectrum. We find the homogeneous core region itself has a negligible impact. The above motivates our choice of large $w$ so that we are able to use analytical formulae for the translucent spectral index and translucent frequency range for a power-law spherical shell (for $p=2$) first derived by \cite{de_bruyn_1976}:

\begin{equation}\label{eq:alpha_sph}
    \alpha_\text{translucent, sph} = \frac{17-17m}{2-8m} ,
\end{equation}

\begin{equation} \label{transl_freq_decades_sph}
    \frac{\nu_1 (r_{\text{min}})}{\nu_\text{br}} = \left(\frac{r_{\text{max}}}{r_{\text{min}}}\right)^\frac{8m-2}{6} .
\end{equation}

\noindent Comparing equation~\eqref{eq:alpha_sph} to equation \eqref{transl_index} we see that for a given $m$, $\alpha_\text{translucent, slab} > \alpha_\text{translucent, sph}$. Comparing equation~\eqref{transl_freq_decades_sph} to equation~\eqref{transl_freq_decades} shows that for an inhomogeneous sphere with equivalent radial dimensions as an inhomogeneous slab, and a given $m$, the translucent part of the spectrum in the spherical case spans a smaller range in frequency. These differences between the slab and sphere geometries will have an effect on the correction factors derived in the next Section.

\subsection{Energy and size implications of a flattened spectrum}\label{sec:results_energy}

\subsubsection{Inhomogeneous slab}\label{sec:inhomog_slab_results}

\begin{figure}
 \includegraphics[width=\columnwidth]{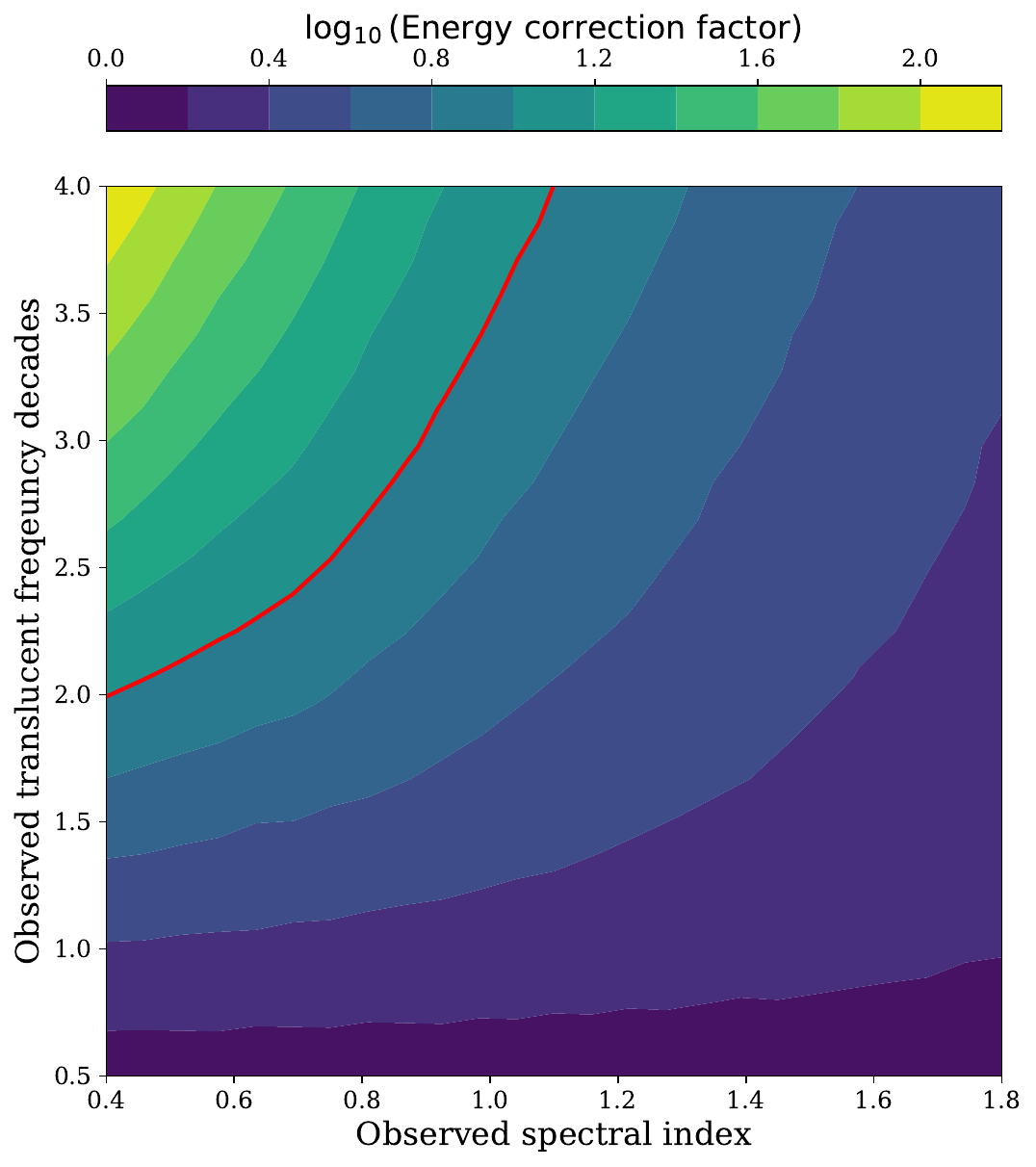}
 \caption{A contour plot of the logarithm to base 10 of the energy correction factor between the equivalent homogeneous model and the inhomogeneous slab model with some translucent spectral index covering the given number of frequency decades. The red line represents energy correction factors of one order of magnitude.}
 \label{fig:energy_ratios_contour}
\end{figure}

\begin{figure}
\includegraphics[width=\columnwidth]{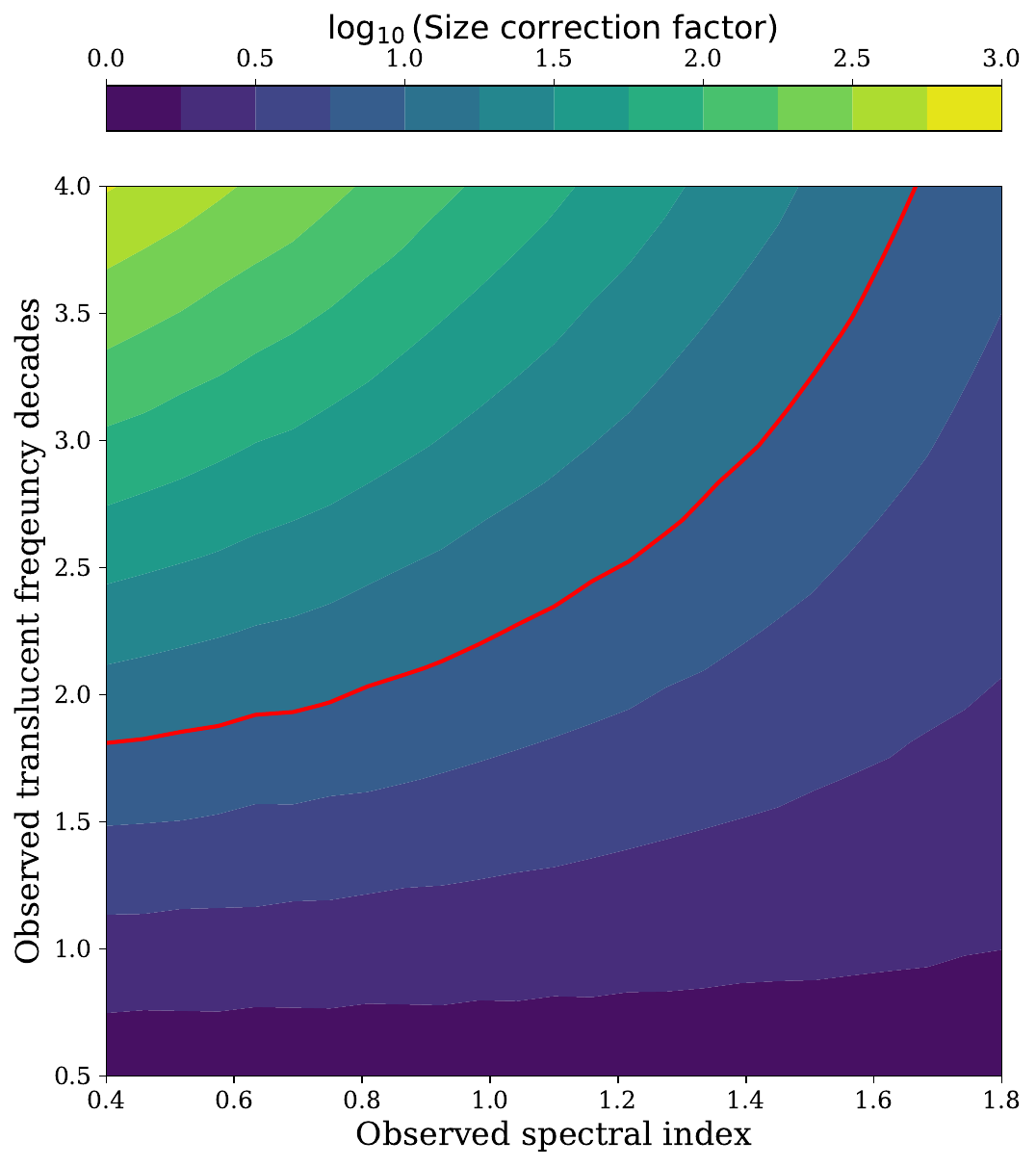}
\caption{A contour plot of the logarithm to base 10 of the size correction factor between the equivalent homogeneous model and the inhomogeneous slab model with some translucent spectral index covering the given number of frequency decades. The red line represents size correction factors of one order of magnitude.}
\label{fig:size_ratios_contour}
\end{figure}

Using both the inhomogeneous slab and sphere models, we can now investigate the impact of using traditional SSA methods, which assume homogeneity, in situations where inhomogeneity is present. We do this by applying the traditional SSA estimates to inhomogeneous models with a range of observable parameters (translucent spectral indices and translucent frequency decades), and calculating the correction factors required to recover the true energy and size from the traditional estimates.

We first construct a regular grid of $\log_{10} \left(\frac{\nu_1 (r_{\text{min}})}{\nu_\text{br}}\right)$ (the translucent frequency decades), and the observed translucent spectral index. The range of observed spectral indices covers the range of flattened indices often seen in observations below the spectral peak, from $0.4$ to $1.8$. The range of observed translucent frequency decades is less easily chosen. We choose a maximum frequency range which could in principle be observed by a dedicated observing campaign utilising observational facilities available today or in the near future. This is discussed further in Section~\ref{sec:discussion}. We then translate these into a series of points of $m$ and $r_{\text{max}}$, for $p=2$, $r_{\text{min}} = 5\times10^8\;\text{cm}$, and $B(r_{\text{min}})=100\;\text{G}$, using equations \eqref{transl_index} and \eqref{transl_freq_decades}. The electron energy distribution cut-offs are taken as Lorentz factors of $10^{0.5}$ and $10^{5}$. We then calculate the inhomogeneous spectrum, the equivalent homogeneous spectrum and the energy and size correction factors for this grid of parameters as outlined in Section~\ref{sec:model}. The specific values of $B(r_{\text{min}})$ and $r_{\text{min}}$ and the electron energy distribution cut-offs do not effect the values of the correction factors. The value of $p$ does have an effect which is discussed later in this Section.

Figure~\ref{fig:energy_ratios_contour} shows a contour plot of the logarithm of the energy correction factor, for the range of observable parameters. To apply this correction factor to an SSA estimated energy from traditional methods, one would identify the observed spectral index below the spectral peak and the number of decades in frequency over which this spectral index is observed before breaking to the $+2.5$ optically thick value. While the observed spectral index below the spectral peak is fairly straightforward to measure (see Section~\ref{sec:motivation} for confounding observational factors), the range in frequency of the translucent part of the spectrum is much more difficult to measure without a wide frequency range of observations below the spectral peak. Methods for estimating the number of translucent frequency decades without needing a large frequency range are discussed further in Section~\ref{sec:discussion}. Overall we see that energy correction factors of over 1 order of magnitude are possible in the range of parameter space where the observed spectral index below the peak is particularly flattened, and/or where this flattened spectral index spans a large range in frequency. This is somewhat intuitive given that these are the cases with the largest unaccounted for bolometric luminosity when compared to the homogeneous model.

Figure~\ref{fig:size_ratios_contour} shows a contour plot of the logarithm of the size correction factor. These size correction factors can then be applied to traditional SSA estimated sizes of emitting regions. We observe that size correction factors of over 2 orders of magnitude are possible over the parameter space, particularly for very flattened spectral indices below the spectral peak, and/or for spectra where the translucent part spans a large range in frequency. This is expected given that the traditional SSA estimates are sensitive only to the peak of the spectrum, so a size estimate close to $r_\text{min}$ is typically obtained, and that having the translucent part span a large range in frequency directly requires a large $\frac{r_\text{max}}{r_\text{min}}$ (from equation \ref{transl_freq_decades}).

\begin{figure}
 \includegraphics[width=\columnwidth]{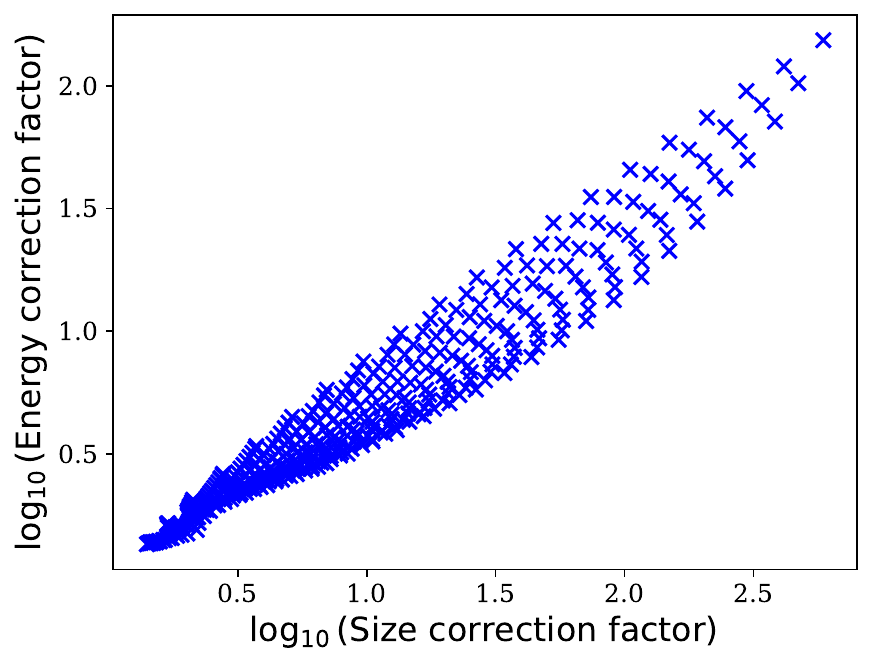}
 \caption{The logarithm to base 10 of the energy correction factor against the size correction factor for the grid of observables discussed in the text.}
 \label{fig:energy_v_size}
\end{figure}

Furthermore, it is not surprising that we observe from Figures~\ref{fig:energy_ratios_contour} and \ref{fig:size_ratios_contour} that the energy and size correction factors are large for roughly the same part of parameter space. Larger emitting regions will (for fixed $r_\text{min}$ and $B(r_\text{min})$) always contain some amount of additional energy. Figure~\ref{fig:energy_v_size} shows the size and energy correction factors for each point in the regular grid described above plotted against each other. We see a clear correlation between the energy and size correction factors as expected, with some scatter. This correlation could for example, be used to estimate the energy correction factor given some \textit{a priori} information about the size of the emitting region, combined with the traditional SSA measurement of the size.

Overall, the results above demonstrate that for reasonable observable parameters, applying traditional SSA estimates to an inhomogeneous source will lead to significant underestimates of both the energy and size of the emitting region. While energy estimates derived from traditional SSA are already usually treated as lower limits, size estimates are seen as robust given the strong dependence of energy on size in the homogeneous case (see equation \ref{eq:enrgy_dependence_size}). However, Figure~\ref{fig:size_ratios_contour} shows that the size estimates from traditional SSA methods \textit{should also be treated as lower limits} if inhomogeneity is present.

We identify two effects which lead to the homogeneous (traditional) SSA estimates requiring correction for the inhomogeneous case. The dominant effect is that in order to generate the flattened spectral index below the peak (e.g. an excess of emission at low frequencies), the inhomogeneous model has an additional large emitting volume, albeit with a lower magnetic field and particle density, compared to the homogeneous model. The second, more nuanced effect, is that due to inhomogeneity, the relationship between the peak frequency of the spectrum, $\nu_\text{peak}$, and $\nu_1$, given by equation \eqref{numax_and_nu1} in the homogeneous case, is altered in the inhomogeneous case. This can be clearly seen in Figure~\ref{fig:spectrum_comparison} where despite having the same peak frequency, the inhomogeneous and homogeneous models have different $\nu_1$ values.

\subsubsection{Magnetic field correction factors}

Traditional SSA estimates also allow for an estimate of the equipartition magnetic field, as outlined in Section~\ref{sec:trad_estimates}. By definition there are a range of magnetic field strengths present in our inhomogeneous slab model. All of these magnetic fields can be thought of as equipartition fields as we have enforced equipartition locally throughout our inhomogeneous slab. We can compare the traditional (e.g. homogeneous) SSA estimate of the equipartition magnetic field with the maximum and minimum magnetic fields present in our inhomogeneous slab. 

We find that the correction factors between traditionally estimated equipartition magnetic fields and the maximum magnetic field values in the inhomogeneous slab are generally small, less than half an order of magnitude over a large parameter space. Furthermore, the traditional SSA estimate can both over and underestimate the maximum magnetic field present in the inhomogeneous case. The calculated correction factors are shown in Appendix~\ref{sec:mag_correction_appendix} in Figure~\ref{fig:max_B_ratios_contour}. These correction factors are small as the region with the maximum magnetic field in the inhomogeneous model dominates the spectrum close to the peak, so traditional SSA estimates perform well. These correction factors arise due to the second effect discussed at the end of Section~\ref{sec:inhomog_slab_results}, the variation of the relationship between $\nu_\text{peak}$ and $\nu_1$ in the inhomogeneous case.

The correction factors between the traditionally estimated equipartition magnetic fields and the minimum magnetic field values in the inhomogeneous slab are strongly dependent only on the observed translucent frequency decades. There is no strong dependence on the observed spectral index below the spectral peak (or equivalently $m$). This is because of the combination of the traditional equipartition method well predicting the magnetic field at $r_\text{min}$ as noted above, and the $m$ dependence in the observed translucent frequency decades (see equation~\ref{transl_freq_decades}). These two results lead to the cancellation of any direct dependence on $m$, and hence on the observed translucent spectral index, of the minimum magnetic field correction factors in the inhomogeneous model. We note that for the inhomogeneous sphere model we find a similar result, where the minimum magnetic field correction factors only have a weak dependence on the observed translucent spectral index.

We find the minimum magnetic field in the inhomogeneous model, $B(r_\text{max})$, is always overestimated by traditional SSA methods, and can be overestimated by up to 3 orders of magnitude over our parameter range. The calculated minimum magnetic field correction factors are shown in Appendix~\ref{sec:mag_correction_appendix} in Figure~\ref{fig:min_B_ratios_contour}. The regions of the inhomogeneous slab which have the minimum magnetic field are those which dominate the spectrum at $\nu_\text{br}$, and the ratio of the minimum to maximum magnetic field is directly proportional to $\frac{r_\text{min}}{r_\text{max}}$ which is in turn proportional to the observed translucent frequency decades (equation \eqref{transl_freq_decades}). Therefore, as we have shown that the maximum magnetic field in the inhomogeneous model is well predicted by traditional SSA methods, the strong dependency of the minimum magnetic field correction factor on the observed translucent frequency decades is expected.

\subsubsection{Variation with $p$}

The results in Section~\ref{sec:inhomog_slab_results} are for $p=2$. We find that varying $p$ over the reasonable range from 2.0 to 4.0 can result in the energy correction factors to vary by up to a factor of $\sim4$ while the size correction factors are considerably less sensitive, varying by less than $\sim30\%$. We find the variation is always such that larger values of $p$ result in larger values of energy and size correction factors.

This result is not surprising as for larger $p$, the value of the optically thin spectral index $\alpha_\text{thin}$ is less (i.e. more negative). Given that the translucent part of the spectrum is emitted by a combination of optically thick and thin regions, a lower value of $\alpha_\text{thin}$ means less total contribution to the flux density from optically thin regions in the translucent part of the spectrum. Therefore, to create a given $\alpha_\text{translucent}$ a greater optically thick contribution is needed, leading to a larger required energy and larger energy correction factor. The correction factors presented in the previous Section are then best thought of as the minimum correction factors over the range of $p$ considered.

\subsubsection{Inhomogeneous sphere}

\begin{figure}
 \includegraphics[width=\columnwidth]{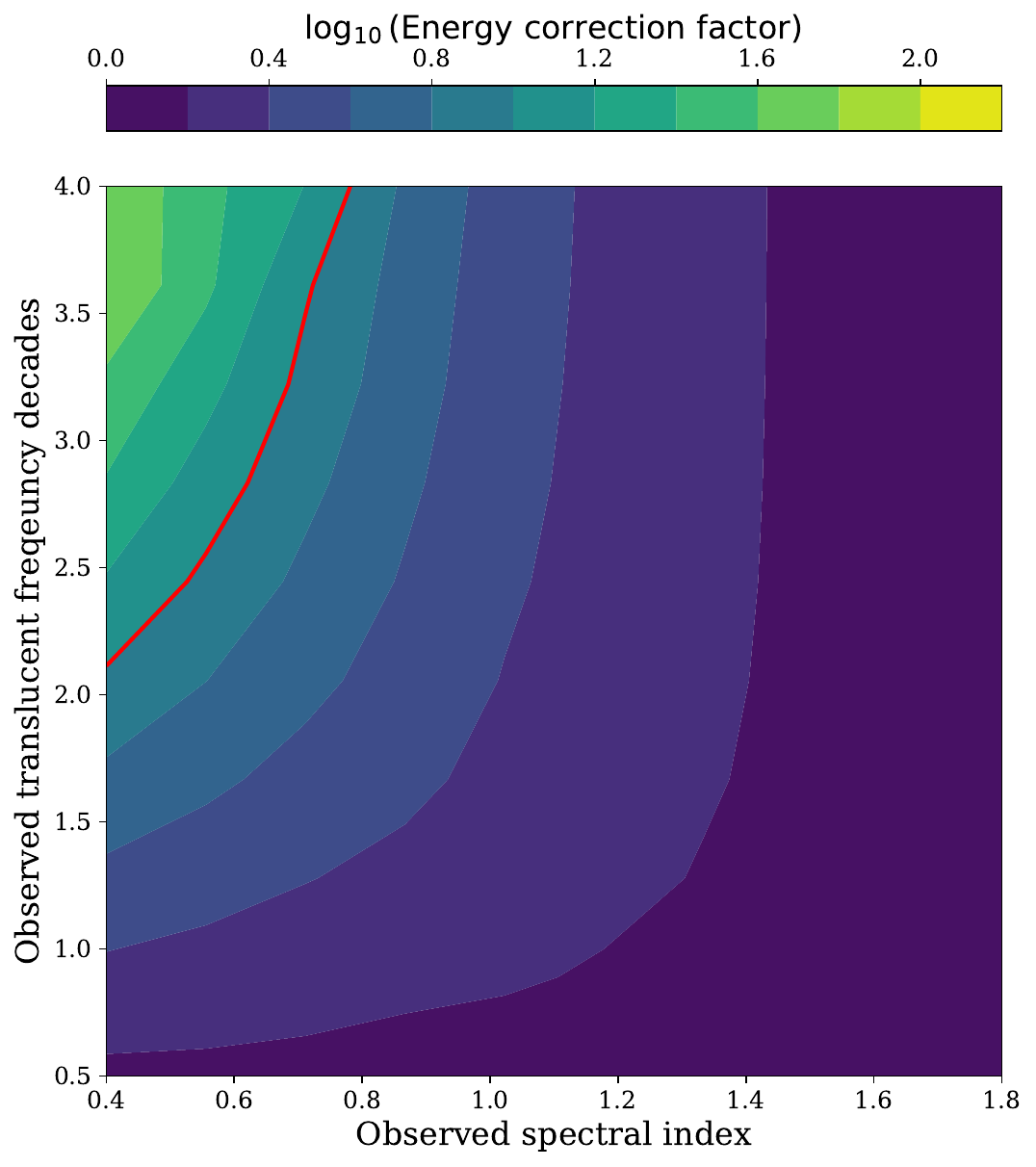}
 \caption{A contour plot of the logarithm to base 10 of the energy correction factor between the inhomogeneous sphere model with the given observational parameters, and the equivalent homogeneous model. The red line represents energy correction factors of one order of magnitude.}
 \label{fig:energy_ratios_contour_sph}
\end{figure}

\begin{figure}
\includegraphics[width=\columnwidth]{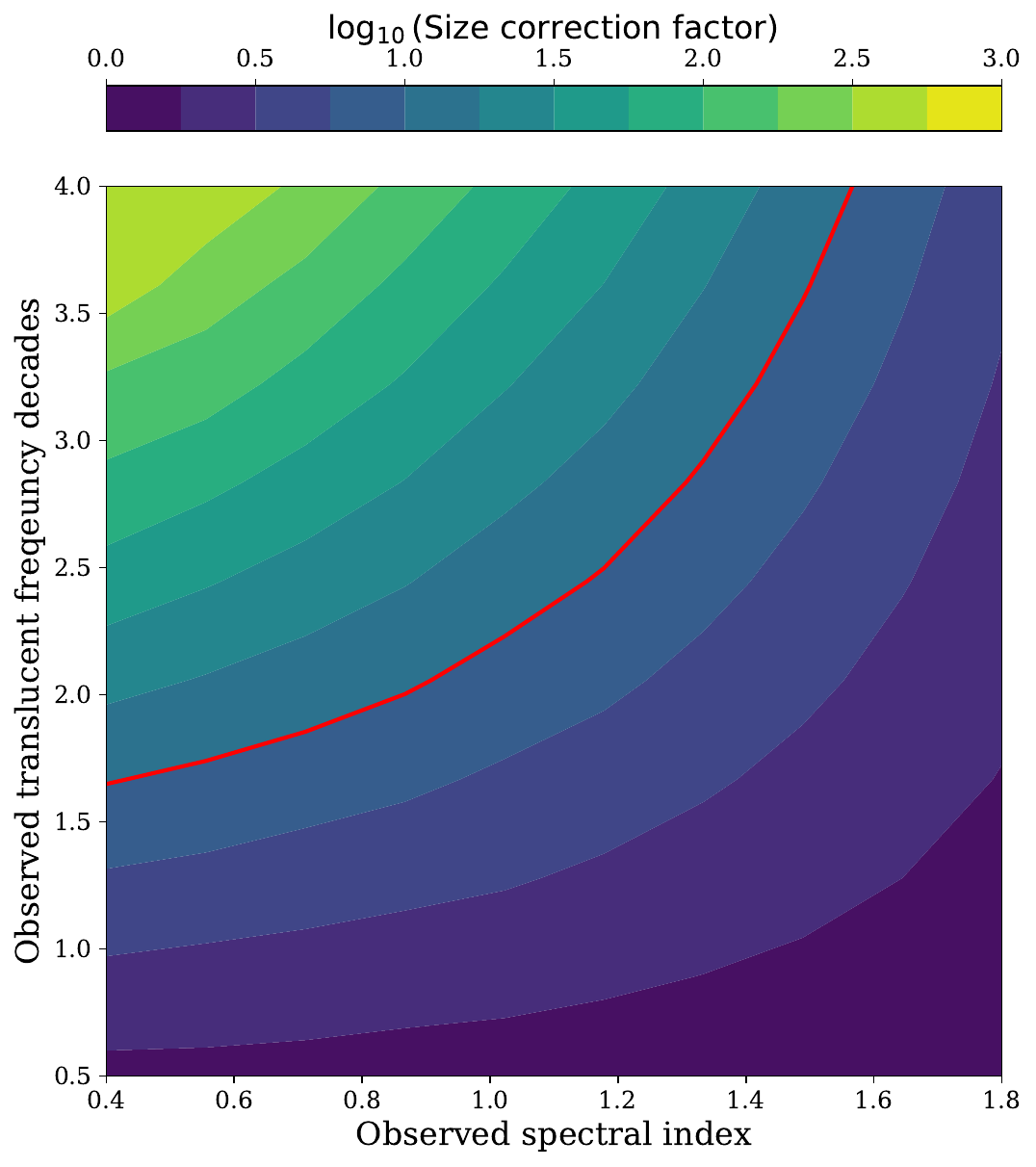}
\caption{A contour plot of the logarithm to base 10 of the size correction factor between the inhomogeneous sphere model with the given observational parameters, and the equivalent homogeneous model. The red line represents size correction factors of one order of magnitude.}
\label{fig:size_ratios_contour_sph}
\end{figure}



We can also investigate the energy and size correction factors for the more realistic model of a power-law sphere with a homogeneous core, described in Section~\ref{sec:model_spherical}. We use the same process and physical values as described at the start of Section~\ref{sec:inhomog_slab_results}. 

Figure~\ref{fig:energy_ratios_contour_sph} shows the ratio between the energy of the inhomogeneous sphere model and the equivalent homogeneous model which has an SSA peak which reproduces the observed spectral peak of the inhomogeneous model - e.g. the energy correction factors. Comparing this to the energy correction factors for the inhomogeneous slab (Figure~\ref{fig:energy_ratios_contour}) we see that the broad trend is similar of higher energy correction factors for lower translucent spectral indices and larger translucent frequency ranges. However, the energy correction factors in the spherical case fall off much quicker with increasing $\alpha_\text{translucent}$. This can be best understood as a consequence of the spherical geometry of the emitting region. For a given observed $\alpha_\text{translucent}$ a larger $m$ is needed for the spherical case over the slab case (see equations \ref{transl_index} and \ref{eq:alpha_sph}) because the outer regions of the sphere have less depth along the line of sight. The larger $m$ leads to the outer regions containing less energy than expected otherwise, lowering the energy correction factor.

Figure~\ref{fig:size_ratios_contour_sph} shows the size correction factors for the inhomogeneous broken power law sphere model. We see that the trend of the size correction factors is similar to the inhomogeneous slab case (Figure~\ref{fig:size_ratios_contour}) and in fact that the size correction factors are in general slightly (up to a factor of $\sim2$) larger for the spherical case.

From this more realistic model we observe that the geometry of the source can have a significant effect on the energy correction factors from inhomogeneity, but that our overall conclusions drawn from the inhomogeneous slab model apply also in this more physical case. Namely, that the presence of reasonable inhomogeneity within a source can cause traditional SSA methods to underestimate the energy and size of the emitting region, in some cases by over an order of magnitude.

Given that we have shown that identifying inhomogeneity and correcting traditional SSA estimates is vital to obtain accurate physical measurements, we can use our simple inhomogeneous slab model to explore other observational signatures of inhomogeneity beyond a flattening of the spectral index below the spectral peak.

\subsection{Polarisation implications of a flattened spectrum}\label{sec:results_pol}

The inhomogeneous slab model explored in this work also allows the opportunity to test other observable features around the spectral peak, and their differences in the homogeneous and inhomogeneous case. Measurement of other observable features can also allow for the more robust identification of inhomogeneity in cases where it is suspected to be present, and allow for its nature to be better understood. One such observable feature is the polarisation properties of the emission around the spectral peak. 

The polarisation properties of synchrotron emission depend strongly on whether the emission is optically thick or optically thin. Equations \eqref{emmission_eq} and \eqref{absorption_eq} give the emission and absorption coefficients for the different polarisations with respect to the magnetic field projection in the plane of the sky. The specific intensity and therefore flux density for each polarisation can be calculated by using the relevant radiative transfer equation. 


Full radiative transfer treatments of polarised synchrotron radiation deal with rotativity (internal Faraday rotation) and conversion, see \cite{jones_odell_1977}. The importance of these effects depends strongly on the minimum energy electron cut-off (or presence of non-relativistic electrons) when compared to the energy of the electrons which have a critical frequency close to the SSA peak. Here we neglect Faraday rotation and conversion as propagation effects and as their full treatment in an inhomogeneous emitting region is beyond the scope of this work.





Making the approximations above, we can simply apply the slab radiative transfer equation, \eqref{intensity_eq}, assuming a uniform magnetic field in direction and strength throughout the homogeneous emitting region. Furthermore, for simplicity we assume that the angle between the uniform magnetic field and the line of sight is $\frac{\pi}{2}$. Altering this direction has no effect on the fractional polarisation or polarisation angle properties in this model, as both polarisations have the same dependence on $\theta$. A magnetic field which is completely uniform in direction throughout the source is unlikely to exist in real astrophysical environments, and any randomness or turbulence in the magnetic field will reduce the observed polarisation fraction. Therefore, the polarisation fractions seen in the model should be taken as upper limits in the perfectly ordered magnetic field limit.

We then calculate the specific intensity for the polarisation perpendicular and parallel to the magnetic field, and then use these to find the fractional linear polarisation, defined as:

\begin{equation}
    |\Pi_\nu| = \frac{|I_{\nu,\perp} - I_{\nu,\myparallel}|}{I_{\nu,\perp} + I_{\nu,\myparallel}} .
\end{equation}

When $\Pi_\nu$ is positive, the observed electric vector position angle (EVPA), or true polarisation angle, is perpendicular to the magnetic field and the and when negative, the observed polarisation is parallel to the magnetic field. 

\begin{figure}
 \includegraphics[width=\columnwidth]{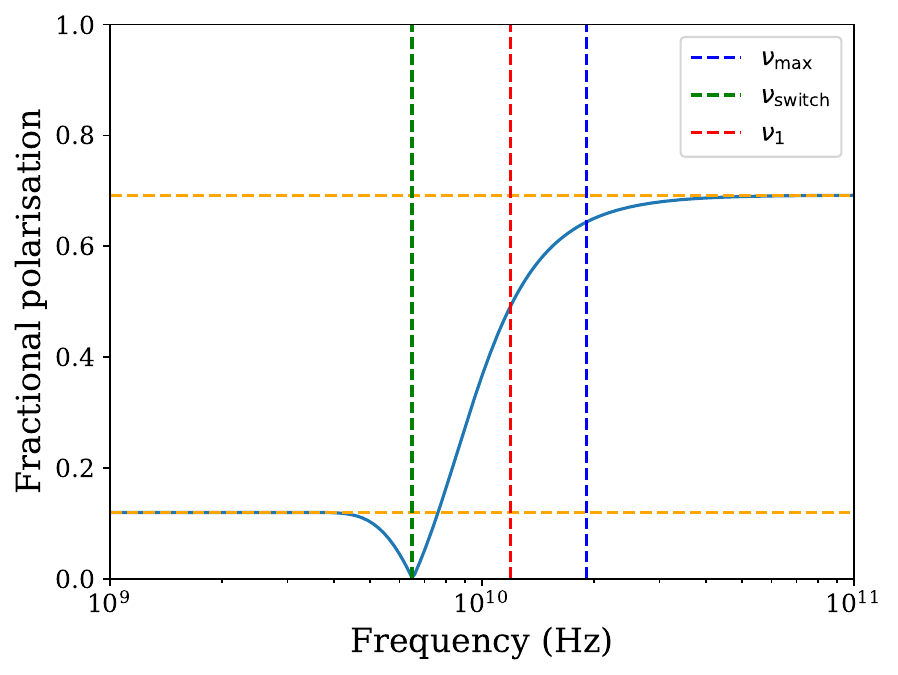}
 \caption{The fractional linear polarisation spectrum for a homogeneous synchrotron source taking into account synchrotron self absorption.}
 \label{fig:homog_frac_pol}
\end{figure}

Figure~\ref{fig:homog_frac_pol} shows the fractional linear polarisation spectrum, $|\Pi_\nu|$, for the same set of parameters used to generate the total intensity spectrum in Figure~\ref{fig:default_sync_spec}. We can see in the optically thin and thick limits the expected linear polarisation fractions of $\sim 70\%$ and $\sim 12\%$ are recovered respectively \citep{pacholczyk}. Furthermore, we see that the observed EVPA changes at $\nu_\text{switch}$ (the quantity $(I_{\nu,\perp} - I_{\nu,\myparallel})$ changes sign), changing from parallel to the magnetic field as expected in the optically thick case, to perpendicular to the magnetic field in the optically thin case \citep{ginzburg_1969}. We note that $\nu_\text{switch} < \nu_1 < \nu_\text{max}$. 


In a similar procedure to the above, we can calculate the fractional linear polarisation spectrum for the inhomogeneous model considered in this work, assuming an ordered magnetic field throughout the emitting region. While many authors have considered robust treatments of certain inhomogeneity and the propagation of polarised radiation (e.g. \citealt{wilson_1980}), here due to the simple slab model used, which is homogeneous along the line of sight, we do not need to consider propagation effects and can instead consider $I_{\nu,\perp}$ and $I_{\nu,\myparallel}$ separately. We use this simple model as a starting point to investigate the linear polarisation of translucent sources, which has not been explicitly explored before. \cite{jones_odell_1977} considered a source inhomogeneous in magnetic field orientation, but homogeneous in all other parameters. \cite{jones_odell_1977_2} considered a homogeneous source with an inhomogeneous boundary. Finally, \cite{curran_2015} briefly consider the polarisation signature from a superposition of several synchrotron sources with differing physical parameters for the case of a flat spectrum source, but no detailed investigations are done.

\begin{figure}
 \includegraphics[width=\columnwidth]{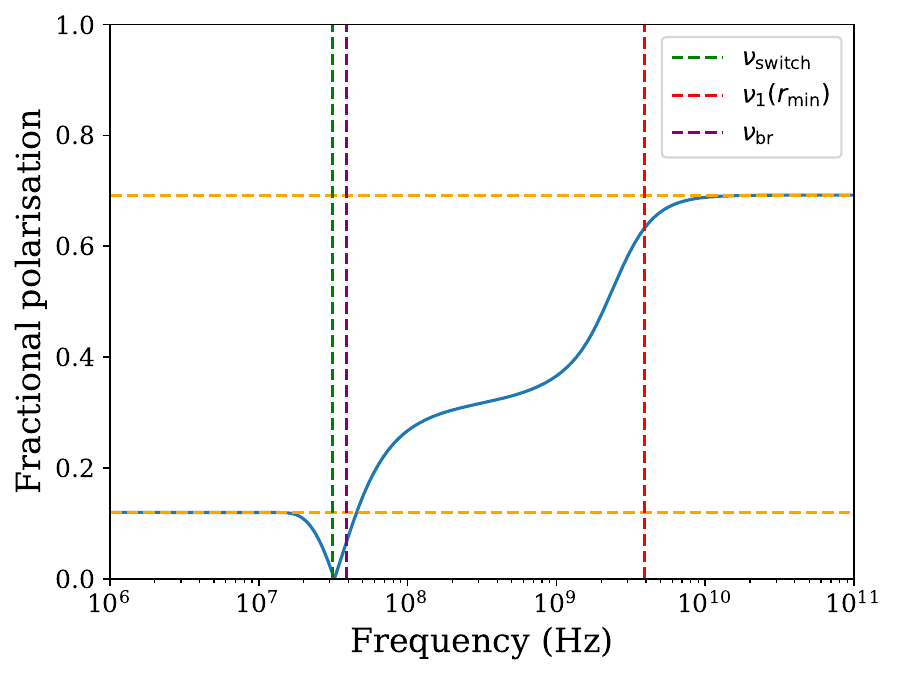}
 \caption{The fractional linear polarisation spectrum for an inhomogeneous source following the model described in Section~\ref{sec:model}, taking into account synchrotron self absorption. The related total intensity spectrum is shown in Figure~\ref{fig:spectrum_comparison} and the parameters used discussed in the text.}
 \label{fig:inhomog_frac_pol}
\end{figure}

Figure~\ref{fig:inhomog_frac_pol} shows the fractional linear polarisation spectrum for the same set of model parameters used to generate the inhomogeneous spectrum in Figure~\ref{fig:spectrum_comparison}. The presence of the translucent part of the spectrum is clearly seen in linear polarisation fraction as well as in total intensity. Once again in the optically thick and thin limits the expected linear polarisation fractions are recovered. However, there are two significant differences between the linear polarisation spectrum in the homogeneous and inhomogeneous case. Firstly, the $\frac{\pi}{2}$ flip expected around the spectral peak in the spectrum is no longer present in the inhomogeneous case, and instead $\nu_\text{switch}$ occurs close to the transition from optically thick to translucent. This occurs because despite the optically thick and thin emission contributing nearly equally in total intensity in the translucent part of the spectrum, the optically thin emission dominates the polarisation throughout the translucent part, and hence no polarisation angle flip is seen close to the spectral peak. Secondly, the level of fractional polarisation at frequencies lower than the spectral peak, is greater than that expected from optically thick emission. This is once again because the optically thin emission dominates, which has a higher fractional polarisation level. 

Numerical testing of our model showed that for any sources with $\alpha_\text{translucent} < 2$ the optically thin emission dominated over the optically thick emission throughout the translucent part of the spectrum. Analytically, considering a single optically thick and single optically thin emission region, a similar result can be derived. \textit{These results mean that any self absorbed synchrotron emission seen with a spectral index of +2 or less may still be dominated by optically thin emission in polarisation.} If this is the case then the EVPA below the SSA will be perpendicular to the projected magnetic field and the level of fractional polarisation may be higher than expected from optically thick emission. Importantly these two properties are expected to be general no matter the model of inhomogeneity used. \cite{jones_odell_1977_2} found a similar result with their treatment of polarised radiative transfer with a homogeneous source and a boundary region along the line of sight, when the boundary region considered contains a significant fraction of the total absorption depth. 

Therefore, as the $+2.5$ spectral index is rarely observed below the spectral peak, if this is due to inhomogeneity or contaminating discrete optically thin emission, then the above results must be taken into account when interpreting polarisation measurements. \cite{curran_2015} pointed this out for the case of the flat spectrum isothermal conical jet model. 

We caveat that while inhomogeneity may be present and identifiable in the total intensity spectrum, in specific cases it may not manifest itself as simply in the polarisation spectrum as demonstrated by our model. For example, if emission volumes with higher magnetic fields also had more significantly more ordered magnetic fields. In this case optically thick emission could still dominate over optically thin emission in the translucent part, and the polarisation flip at the spectral peak would still be observed. This may be expected for synchrotron emission from particles accelerated at shock fronts which order and amplify magnetic field (\citealt{laing_1980}, \citealt{bell_2004}). Additionally, other factors can lead to confounding changes in polarisation angle. For example, changes in magnetic field geometry over time or bulk relativistic motion differences throughout the emission region (e.g. \citealt{lyutikov_2005}). 


The results above mean that observations of the fractional linear polarisation spectrum around the spectral peak have the potential to unambiguously identify inhomogeneity in a synchrotron source, and provide additional information about the magnetic field ordering and geometry in the emitting region. Observations at many frequencies around the spectral peak are crucial for this. While simultaneous polarimetric observations at multiple frequencies can accomplish this, for many sources of interest the spectral peak is observed as the emitting region expands and transitions from optically thick to thin. For these synchrotron transients, observing the time dependence of the polarisation as the spectral peak moves in frequency presents a unique opportunity, although certain conflicting effects, such as magnetic field evolution, may also play a role. In the radio regime, short timescale polarisation monitoring of X-ray binary systems during transitions from optically thick to thin is a promising possibility for identifying these effects (\citealt{hughes_2023}, Cowie et al. in prep.).

\subsection{Time dependent properties of a flattened spectrum}\label{sec:time}

SSA is often observed in synchrotron emitting transients which evolve with time. The qualitative model for these transients is a synchrotron emitting region which expands over time. Under a broad range of parameters, this expansion will result in the emitting region becoming more optically thin with time. Observations at a single frequency will then show a rise in flux while the source remains optically thick as the area of the source increases, followed by a peak which corresponds to the transition from optically thick to optically thin, the SSA peak, and a decline as the source continues to lose energy as it expands. This model was outlined by \cite{vdL}, with particular focus on a model with adiabatic expansion and conservation of magnetic flux. This model is still widely used to interpret flares from synchrotron transients where the change from optically thick to thin is seen (e.g. \citealt{zadeh_2006}, \citealt{tetarenko_2017}). However, this simple model does not explain several features of some observations (e.g. \citealt{fender_2023}), and comparison to hydrodynamic simulations where a bulk velocity is present show that an expanding homogeneous quasi-spherical emitting region does not survive for long \citep{savard_2025}. Therefore, here we explore an extension to this time dependent model where we introduce inhomogeneity of the synchrotron emitting region.

The introduction of inhomogeneity allows for far more complex models to be investigated when compared to the one presented in \cite{vdL}. Here we focus on a simple quasi-spherical model where the degree of inhomogeneity remains constant in time, in order to investigate the general implications of inhomogeneity in the time domain. We stress that more complex models have the potential to exhibit different properties, but investigating these is beyond the scope of this work. We take the inhomogeneous quasi-spherical cylindrical slab model presented in Section~\ref{sec:model} and allow, $B(r_{\text{min}})$, $N_0(r_{\text{min}})$, $r_{\text{min}}$ and $r_{\text{max}}$ to vary with time. To find how each of these vary with time we make the following assumptions. We assume that the source expands at some constant velocity, $v$, such that $r_{\text{max}}(t) = vt + r_{\text{max}}(0)$. We require that the depth evolve as $s=\frac{4}{3} (r_{\text{max}}(t)^2 - r_{\text{min}}(t)^2)^\frac{1}{2}$, so that the emitting region remains quasi-spherical. We assume that the degree of inhomogeneity throughout the source remains constant with time, such that $\frac{B(r_{\text{min}})}{B(r_{\text{max}})}$ is constant in time. Therefore, $r_{\text{min}}(t) = r_{\text{max}}(t) \frac{r_{\text{min}}(0)}{r_{\text{max}}(0)}$ and the expansion is self similar. We assume a constant ratio between the energy in the magnetic field and electrons in time, in this case choosing a ratio so that the model is close to equipartition. Finally, we assume that the relativistic plasma cools adiabatically so that for a single particle $E \propto V^{-\frac{1}{3}}$ should hold. Then since $V \propto r_{\text{max}}^3$ we obtain $E \propto r_{\text{max}}^{-1}$. Finally we assume that particle number is conserved (see equation 4d of \cite{vdL}) which leads us to $N_0 (r_{\text{min}}) \propto t^{-(p+2)}$. These conditions lead to the same evolution of $N_0(r_\text{min})$ and $B(r_\text{min})$ as in the model presented in \cite{vdL}.

\begin{figure}
 \includegraphics[width=\columnwidth]{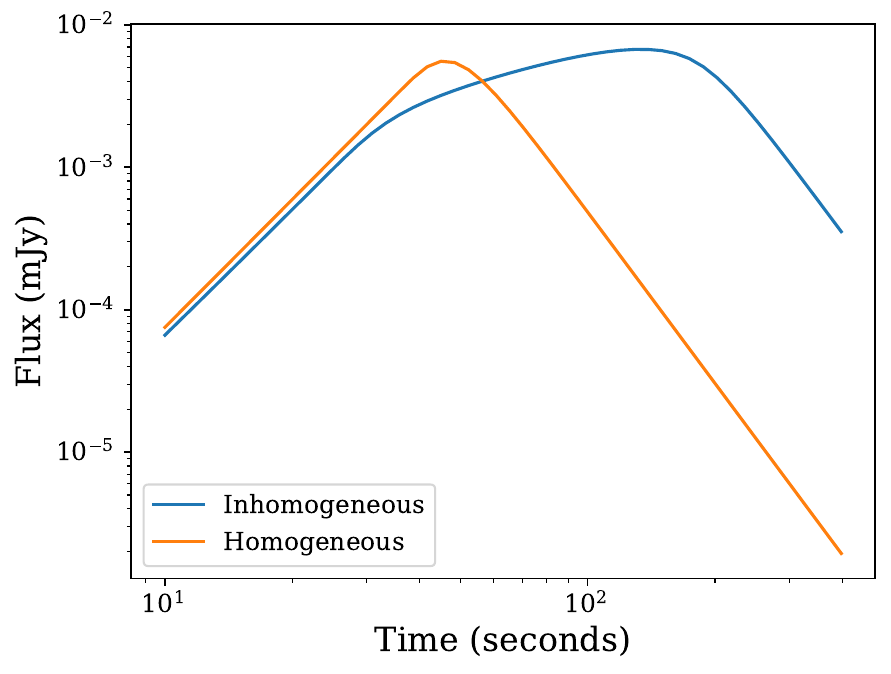}
 \caption{Lightcurve of a synchrotron emitting inhomogeneous region expanding at $0.1c$ observed at 1.3~GHz, and a comparison to the equivalent homogeneous model expanding at the same speed. The full model and the parameters used are detailed in text.}
 \label{fig:lightcurve}
\end{figure}

Armed with the conditions for how each variable in our expanding inhomogeneous slab model should evolve with time, we can investigate what a lightcurve at a single frequency looks like, and compare it to the homogeneous case \citep{vdL}. We choose $v=0.1c$, $r_{\text{max}}(0)=5\times10^8 \: \text{cm}$, and $\frac{r_{\text{max}}}{r_{\text{min}}} = 5$, $B(r_{\text{min}},0)=1\times10^5 \: \text{G}$ and $m=3$. We choose the comparison model as the homogeneous model which produces an equivalent spectrum at $t=0$ and evolve this model with the same velocity, according to the conditions set out in \cite{vdL}. Figure~\ref{fig:lightcurve} shows the lightcurves from the inhomogeneous and homogeneous model, in this case at a observing frequency of 1.3~GHz. The homogeneous model shows the characteristic power law evolution in time as expected by the model presented in \cite{vdL}. This can be derived by characterising the optically thick and thin regions of the spectrum as a power law: $F_\nu = F_{\nu_1} \left(\frac{\nu}{\nu_1}\right)^\alpha$. Then by using the time evolution of $B$, $N_0$ and $R$ and equations \eqref{nu1_eqn} and \eqref{fnu1_eqn}, we can show:

\begin{equation}\label{eqn:time_coeff}
    \beta = \frac{7}{3}\alpha - \frac{17}{6} ,
\end{equation}

where $F_\nu \propto t^\beta$. Then the optically thick rising portion of the lightcurve has $\beta = 3$ and the optically thin falling portion of the lightcurve has $\beta = -4$. 

The equation above holds for any region of the spectrum which can be characterised as a power law, so long as the relevant physical quantities have the same time evolution and the power law spectrum is constant in time. Therefore, for the inhomogeneous model where the degree of inhomogeneity is constant in time, we can use equation \eqref{eqn:time_coeff} and the value of $\alpha_\text{translucent}$ to predict the behaviour of the lightcurve when the observing frequency lies in the translucent part of the spectrum. We find that for $\alpha_\text{translucent} > 0$ we have $-\frac{17}{6} < \beta_\text{translucent} < 3$. 

Therefore, the qualitative difference between the homogeneous and inhomogeneous models is that the steep time dependence in the rising portion of the lightcurve in the homogeneous case can be flattened out or even inverted in the inhomogeneous model. Furthermore, in the case where $\beta_\text{translucent} > 0$ the equivalent homogeneous model peaks in time before the inhomogeneous model. This is because the peak of the spectrum in the inhomogeneous model is dominated by a small, dense, highly magnetised volume, which takes more time to transition to becoming optically thin when compared to the equivalent homogeneous model. 



When $\beta_\text{translucent} < 0$ ($\alpha_\text{translucent} < \frac{17}{14}$), the peak observed in the lightcurve does \textbf{not} correspond to the peak in the spectrum. Therefore, when using time resolved observations at a single frequency to identify the spectral peak, care must be taken where inhomogeneity may be present, as the peak in the lightcurve may not correspond to the actual spectral peak. Instead the spectrum will remain translucent for a time after the peak in the lightcurve, until the transition to optically thin is made. Finally, we note that other predictions from the model presented in \cite{vdL}, such as the ratio of peak fluxes at two different observing frequencies, are unaffected by inhomogeneity under the conditions assumed.

We stress that the model of time evolution considered to here is specific, both in how the inhomogeneity evolves and in using the same assumptions used by \cite{vdL}. Several possible scenarios may change the observed lightcurve, for example, extended periods of particle acceleration, evolution of the inhomogeneity, and non-spherical expansion. These extensions to this model will be explored in future work, as there remain many observable features of SSA radio flares  that are not satisfactorily explained by the model above. However, while the particular prediction relating $\beta_\text{translucent}$ to the observed $\alpha_\text{translucent}$ is specific to this model, the general effect of inhomogeneity is apparent. Inhomogeneity in an expanding emitting region will result in a slower than expected increase/decrease in flux when compared to a homogeneous model, broadening the peak of the lightcurve. Inhomogeneity therefore also results in a delayed peak in the lightcurve when compared to equivalent homogeneous models with the same expansion velocity. This means for events where there is no \textit{a priori} knowledge of the start time of the flare, applying a homogeneous model will result in an underestimate in both time elapsed since the start of the flare, and the size of the emitting region when the peak of the lightcurve is reached at the observing frequency (see Section~\ref{sec:results_energy}).



\section{Discussion}\label{sec:discussion}


\subsection{Improved SSA energy estimates}

In order to obtain the most accurate minimum energy, size and equipartition magnetic field estimates from observations of the SSA peak, whether from measurement of a turnover in the spectrum, or a peak in an lightcurve, one can start either from equations \eqref{eq:energy_homog_nu_form}-\eqref{eq:B_homog_nu_form} if the synchrotron cut-off frequencies can be constrained; or from equations \eqref{eq:e_homog_E}-\eqref{B_homog_E} if the electron energy cut-offs can be constrained. Even if there are large uncertainties on the cut-off energies or frequencies, the equations in Section~\ref{sec:trad_estimates} should still be used in order to understand how these uncertainties impact the respective estimates of physical parameters. However, this traditional SSA method assumes a quasi-spherical and homogeneous emitting region, not likely in real astrophysical environments, and as we show in Sections~\ref{sec:results_spehrical} and \ref{sec:results_energy} respectively, using these traditional methods can result in significant errors when these effects are present. Therefore, to move towards more accurate measurements of physical parameters using SSA, correction factors for these effects should be considered.


In Section~\ref{sec:results_spehrical} we showed that for homogeneous emitting regions which are non-spherical both over and underestimates of parameters such as the size and minimum energy are possible. If the ratio of the depth to the observed area of the emitting region is known then these errors in the traditional SSA estimate can be corrected using the formalism of equations \eqref{E_with_sphericity}-\eqref{V_with_sphericity}. Unfortunately, only in rare cases is this ratio known. However, calculating the correction factors for a physically reasonable range of sphericities will give insight into the uncertainty on parameters estimated using traditional SSA.

In Section~\ref{sec:motivation} and Section~\ref{sec:results_energy} we demonstrated that when a flattened spectral index below the spectral peak is observed then inhomogeneity is a likely physical interpretation. In these cases when applying the traditional SSA parameter estimates the resulting measurements of the size and minimum energy can be underestimated by over an order of magnitude. To more accurately recover the physical parameters, the traditional SSA parameter estimates should be corrected using the correction factors derived in Section~\ref{sec:results_energy}. To calculate the correction factors the spectral index at frequencies below the peak must be measured. Furthermore, a measurement of the frequency range over which this flattened spectral index holds (the translucent part of the spectrum) is needed. However, there are two potential complications to this method. 

The first is that while we have verified the correction factors to be independent of absolute values of size and magnetic field etc. (see Section~\ref{sec:model_slab} for a full discussion), we find that the minimum energy correction factors specifically are sensitive to the geometry of the emitting region (Figure~\ref{fig:energy_ratios_contour} compared to Figure~\ref{fig:energy_ratios_contour_sph}). Furthermore, the models explored in this work are deliberately simple parametrisations of inhomogeneity, and one can imagine that for true astrophysical sources the situation is more complex. Nevertheless, the inhomogeneous models used, and the correction factors derived represent a step towards improved measurements using observations of SSA. The correction factors for the minimum energy can therefore be applied to observations where the geometry is not known, so long as the appropriate uncertainty is acknowledged. We find that the size correction factors are much less sensitive to the emitting region geometry (Figure~\ref{fig:size_ratios_contour} compared to Figure~\ref{fig:size_ratios_contour_sph}). Therefore, despite not knowing the geometry or specific parametrisation of the inhomogeneity for a source, corrections to size estimates from traditional SSA methods can be made with more confidence.

The second complication to applying the correction factors we have derived is that it is typically difficult to constrain the range of frequencies which the translucent part of the spectrum spans, necessary to calculate the correction factors. This can require observations over several orders of magnitude in frequency, currently only possible with a dedicated (simultaneous in the case of transients) observing campaign involving several facilities. Currently, in the most optimistic scenario 3-4 orders of magnitude in frequency could be covered by existing instruments which observe synchrotron radiation (e.g. the LOFAR interferometer \citep{lofar}, to the James Clerk Maxwell Telescope). In most scenarios a lower limit on the number of translucent frequency decades will be measured, and hence only a lower limit on the minimum energy and the size correction factors will be obtained. This can and should still be used to correct traditional SSA estimates.

In some cases, particularly those of transients or time evolving sources, an upper limit on the size may be available when considering the expansion speed of the region. Applying traditional SSA methods to obtain a size estimate, combined with an estimate of the flare/event start time, allows for an estimate of the expansion speed. Then with the assumption that this cannot be larger than $c$ (only the case when not taking into account the relativistic Doppler effect), a tentative maximum size correction factor can be obtained. This, when combined with the measurement of $\alpha_\text{translucent}$, can be used to estimate the range in frequency over which the translucent part of the spectrum covers, and using Figure~\ref{fig:energy_v_size} can provide an estimate of the relevant energy correction factor needed. This allows for inhomogeneity to be accounted for, without needing observations over several orders of magnitude in frequency. Other similar arguments may also be possible from energetic considerations, hence why Figure~\ref{fig:energy_v_size} is provided.

Overall, there are existing observational datasets (and no doubt more to be taken in the future) to which traditional SSA methods are applied. The measurements of minimum energy, emitting region size and equipartition magnetic field from these observations can be significantly improved by applying the formalism and correction factors outlined in this work. 




\subsection{Inhomogeneity in astrophysical sources}

In Section~\ref{sec:motivation} we showed that there are different situations which can lead to a spectral index below the spectral peak which is less than the expected $+2.5$. We argue that with good quality observational data around the spectral peak, spanning around an order of magnitude in frequency and ideally using a combination of inter and intra-band spectral indices where available, it is possible to discern between these different situations. We further assert that in many cases, with the current observational evidence, in a variety of astrophysical synchrotron sources, inhomogeneity seems to be the likely explanation for the observed flatter spectral indices. 

For example, one clear observational example where no other mechanism but inhomogeneity can be sensibly invoked to explain the spectrum is for SN 2019oys at late times. In this case the spectrum presented in Figure~5 of \cite{sfaradi_2024} shows a spectral index below the peak which is clearly flatter than $+2.5$. Traditionally these radio loud SNR observations were fit with a phenomenological model where the effect of inhomogeneity was introduced via reducing the sharpness of the spectral break between the optically thick and thin regions of the spectrum \citep{soderberg_2005}. While the qualitative consequences of this were recognised by \cite{bjornsson_2017} (including broadening of the lightcurve peak as predicted by our time dependent model in Section~\ref{sec:time}), our analysis can be used to estimate the quantitative correction factors for SNe where flatter spectral indices are observed below the spectral peak. The physical cause of the inhomogeneous synchrotron emitting regions in SNe is thought to be inhomogeneities in the circumstellar medium which when shocked produces synchrotron emission \citep{sfaradi_2024}. Other factors such as inhomogeneity in the SNe ejecta, and time evolution of the shock front and the related particle acceleration may also have similar effects. Many SNe observations at radio wavelengths show signatures of this inhomogeneity (e.g. \citealt{soderberg_2005}, \citealt{weiler_2011}, \citealt{corsi_2014}, \citealt{bjornsson_2017}), and are prime examples of where the correction factors discussed in this work could be applied.

However, inhomogeneity is, at the time of writing, rarely discussed in the context of other astrophysical synchrotron emitters. For example in the case of GRB afterglows, better observational data, particularly at radio wavelengths, is showing that in some cases more complex models, with multiple synchrotron emitting components, are needed to fit the data (e.g. \citealt{rhodes_2024}). Inhomogeneity within one or two components could also potentially explain these observations. To our knowledge the only consideration of inhomogeneity in GRB afterglows is the specific case of fast cooling of electrons, explored by \cite{granot_2000}. They find that this specific model of inhomogeneity flattens the spectral index to $+\frac{11}{8}$. More general inhomogeneity can give a range of possible spectral indices as shown in this work, and could be present in both the forward shock and reverse shock synchrotron emitting components thought to make up the GRB afterglow. In the case of the forward shock, the ejecta from any GRBs are expected to propagate into a CSM created by the stellar wind of the precursor, which similarly to the SNe cases discussed above, may contain significant inhomogeneity. In the case of the reverse shock, if the GRB jet is has structure within it, as some recent results evidence (e.g. \citealt{cunningham_2020}), then there will be inhomogeneity in the jet and hence reverse shock emission region. Observations of the spectral index of the self absorbed reverse shock emission can then probe the structure in the jet, and as illustrated in this work, provide more accurate energetic estimates. However, we note that, flattened spectral indices below the spectral peak are fairly uncommon in GRBs when compared to other source classes such as XRBs, SNe, FBOTs, and TDEs. One explanation for this observation is that significant relativistic beaming in the GRB case causes only a small (homogeneous) part of the emitting region to dominate the spectrum, meaning any wider inhomogeneity is not observed.

Inhomogeneity is also not considered in another class of synchrotron transients, XRBs. Optically thick to thin synchrotron flares are seen in XRBs on timescales of minutes to hours around the time of the state transition from the hard state to the soft state \citep{fender_2004}. Although these flares are likely somehow linked to the launch of the discrete jet ejecta \citep{wood_2021}, the nature of the flaring region remains unknown. However, these synchrotron flares provide an opportunity to quantify the energy these systems release, invaluable for using these systems as time resolved jet and accretion laboratories. 

These synchrotron flares are commonly modelled using the homogeneous model presented in \cite{vdL}, as discussed in Section~\ref{sec:time}. The energy of the plasma responsible for these flares is then simply estimated by applying traditional SSA methods to the peak in the lightcurve. Then expansion energetics as well as expansion speeds and  for the emitting region can be found, by utilising the start time of the flare when known. Typical expansion speeds found are on the order of $0.1c$ \citep{fender_bright_2019}. However, as in many other sources the $+2.5$ spectral index is rarely seen in XRBs. If this is due to inhomogeneity then corrections to both the energy of the flare and the expansion speed of the emitting region are needed. Furthermore, direct evidence of inhomogeneity from resolved imaging has been found at times after the flare in the discrete ejecta from XRBs (e.g. \citealt{rushton_2017}).

In addition to the different source classes discussed above, SSA peaks are observed in more exotic synchrotron transients, such as FBOTs and TDEs, and in both cases spectral indices deviating from the expected $+2.5$ are seen (e.g. \citealt{stein_2021}, \citealt{nayana_2025}). While there is limited observational data on these relatively new source classes at the moment, as higher quality data becomes more common it will become possible to unambiguously identify inhomogeneity and correct for its effect.

It perhaps is not surprising that inhomogeneity is likely widespread across different astrophysical sources, as homogeneity is a rather unique condition that can be easily broken by a number of physical mechanisms. Therefore, in cases where a $+2.5$ (or $+2$ in the case of $\nu_{cr} (\gamma_\text{min}) > \nu_1$) is confidently measured, what does this illustrate? Unfortunately, while this observation could demonstrate a truly homogeneous emitting region, it is also possible for certain parametrisations of inhomogeneity to be present in the emitting region and \textit{not} effect the synchrotron spectrum. This "hidden" inhomogeneity can occur in our slab model for example when $m<1$, but will also likely exist in more general models. However, we observe in the slab model that situations where there is hidden inhomogeneity typically have very low energy and size correction factors, and so traditional SSA methods are accurate in these cases.

\section{Conclusions}\label{sec:conclusion}

In this work we have used models of synchrotron emitting regions to investigate how inhomogeneity affects measurements of energy, size, magnetic field and polarisation using the SSA peak. Traditional SSA methods to estimate physical parameters from measurements of the SSA peak assume both homogeneity and quasi-sphericity, and we investigate the effect of deviations from both of these assumptions. Numerous conclusions can be drawn from our work:

\begin{itemize}
    \item{In Section~\ref{sec:homog} (and Appendix~\ref{sec:full_equations}) we derived the most accurate set of SSA energy estimate equations assuming a homogeneous model available to date, with full dependence on the details of the power-law electron energy distribution included. These are suitable to apply to a wide variety of observations. Furthermore, we provide clear correction factors for cases where the geometry is not quasi-spherical, and find that underestimates and overestimates of all parameters are possible depending on the geometry.}
    \item In Section~\ref{sec:motivation} we argued that inhomogeneity is the most likely explanation for flattened spectral indices below the spectral peak in a variety of astrophysical scenarios.
    \item{In Section~\ref{sec:results_energy} we for the first time have derived energy and size correction factors to traditional SSA estimates which assume homogeneous emitting regions, for the case where inhomogeneity is present. We showed that the presence of a flattened spectral index below the spectral peak, if due to inhomogeneity, can result in traditional SSA methods underestimating the energy and size by over an order of magnitude. Using both a spherical and slab model of the emitting region we demonstrated that the energy correction factors can depend on the geometry, but that the size correction factors are substantially less sensitive to this.}
    \item{In Section~\ref{sec:results_pol} we investigate how inhomogeneity changes the polarisation properties around the spectral peak, and show that there exist measurable differences between the inhomogeneous and homogeneous fractional linear polarisation spectra, which may allow for further identification and characterisation of inhomogeneity beyond using the total intensity spectrum.}
    \item{In Section~\ref{sec:time} we explore a simple time dependent model of an expanding inhomogeneous region, and compare the lightcurve this generates to the homogeneous model. We explore the qualitative differences, and illustrate that the main effect is flattening of the steep rise in the lightcurve when inhomogeneity is present.}
    \item{Finally, in Section~\ref{sec:discussion} we discuss how to practically apply the correction factors to observations and briefly explore the implications of taking inhomogeneity into account for different source classes.}
    
\end{itemize}

Overall, we have used models of inhomogeneity in a synchrotron emitting region to demonstrate the various effects introduced when compared to a homogeneous region. The quality of spectroscopic and polarimetric observations in the radio regime is reaching a stage where the presence of inhomogeneity is apparent, and it must be taken into account in order to obtain more accurate estimates of the physical conditions in, and the energetics of, various astrophysical sources. The formalism outlined in this work provides the necessary steps towards improving our energy and size estimates using SSA observations by and can be directly applied to current observations of a wide variety of astrophysical sources.

\section*{Acknowledgements}

FJC was supported by STFC grant ST/Y509474/1. RF acknowledges support from UK Research and Innovation, the European Research Council and the Hintze Charitable Foundation.

\section*{Data Availability}

The code used for this work is publicly available at \url{https://github.com/Fraserjcowie/SSA_estimates}.



\bibliographystyle{mnras}
\bibliography{flatspectrabib} 




\appendix

\section{Full equations and derived quantities}\label{sec:full_equations}

In this appendix we list the constants and full set of equations in both the energy cut-off and frequency cut-off formalisms for calculating the energy, magnetic field and size of a homogeneous quasi-spherical emitting region from observations of the SSA peak. Full details on treatments of the different formalisms, inclusion of non-thermal protons and deviation from equipartition, and redshift and Doppler factor corrections, see the further Sections of this appendix. In addition, we give the full equations for derived quantities such as the total number of non-thermal particles and the non-thermal particle density.

Throughout we use constants and pseudo-constants weakly dependent on the electron energy distribution index, $p$, first defined in \cite{pacholczyk}:

\begin{equation}
    c_1 = \frac{3e}{4\pi m_e^3c^5} ,
\end{equation}

\begin{equation}
    c_5(p) = \frac{\sqrt{3} e^3}{16 \pi m_e c^2} \Gamma\left(\frac{3p-1}{12}\right) \Gamma\left(\frac{3p+7}{12}\right) \Gamma\left(\frac{p+\frac{7}{3}}{p+1}\right) ,
\end{equation}

\begin{equation}
    c_6(p) = \left(\frac{c}{c_1}\right)^2\frac{\sqrt{3} e^3}{128 \pi m_e c^2} \left(p+\frac{10}{3}\right)\Gamma\left(\frac{3p+2}{12}\right) \Gamma\left(\frac{3p+10}{12}\right) ,
\end{equation}

\noindent where $\Gamma$ is the ordinary gamma function. We use these to define two further pseudo-constants $k_1$ and $k_3$, where the form of $k_3$ depends on whether the frequency cut-off ($k_{3\text{,}\nu}$) or energy cut-off ($k_{3,E}$) formalism is used:

\begin{equation}
    k_1(p) = \left(\frac{\pi c_5(p)}{c_6(p)}\right)^2 (2c_1)^{-5} (1-e^{-1})^2 ,
\end{equation}

\begin{equation}
    k_{3\text{,}\nu}(p) = \left(2 c_1\right)^{-\frac{p+4}{34}} \left(\frac{1}{8\pi} \frac{4}{3} \frac{11}{6} c_6(p)\right)^{-\frac{1}{17}} ,
\end{equation}

\begin{equation}
    k_{3,E}(p) =\left( \frac{11}{2(p+1)} \frac{1}{8 \pi} \frac{4}{3} c_6(p)\right)^{-\frac{1}{1+2(p+6)}} (2c_1)^{-\frac{p+4}{2+4(p+6)}} .
\end{equation}

\noindent In addition, we define the variables $K_\nu$ and $K_E$ which contain information on the electron energy distribution in the two different formalisms:

\begin{equation}
    K_\nu = \int^{\nu_\text{max}}_{\nu_\text{min}} \frac{\nu^{-\frac{p}{2}}}{2} c_1^\frac{p-2}{2} d\nu ,
\end{equation}

\begin{equation}
    K_E = \int^{E_\text{max}}_{E_\text{min}} E^{1-p} dE ,
\end{equation}

\noindent Where $E_\text{max}$, $E_\text{min}$ and $\nu_\text{max}$, $\nu_\text{min}$ are the high and low energy cut-offs in the electron energy distribution or frequency cut-offs in the synchrotron spectrum.

Finally, the fraction of energy in non-thermal protons compared to non-thermal electrons is $\eta$ (Appendix~\ref{sec:equipartition_deviation}), the deviation from equipartition is parametrised with $\epsilon_{\nu}$ or $\epsilon_E$ (Appendix~\ref{sec:equipartition_deviation}), and the Doppler factor, cosmological redshift and luminosity distance of the emitting region are $\delta$, $z$, and $D_L$ respectively (see Appendix~\ref{sec:doppler_and_redshift}). The full equations in the frequency cut-off formalism, where $F_{\nu_1}$ and $\nu_1$ are the \textit{observer frame measured quantities}, are then:

\begin{align}\label{eq:final_energy_nu_form}
    E_\text{homog} = & \frac{17}{36} k_1^{-\frac{10}{17}} k_{3\text{,}\nu}^{11} K_\nu^\frac{11}{17} F_{\nu_1}^\frac{20}{17} D_L^\frac{40}{17} \nu_1^\frac{11p-56}{34} \delta^{-\frac{64+11p}{34}} (1+z)^{\frac{11p-96}{34}} \nonumber \\ & (1+\eta)^{\frac{11}{17}}  \left(\frac{6}{17}\epsilon_\nu^\frac{11}{17} + \frac{11}{17}\epsilon_\nu^{-\frac{6}{17}}\right) ,
\end{align}

\begin{align}\label{eq:final_r_nu_form}
    R_\text{homog} = & k_1^{-\frac{4}{17}} k_{3\text{,}\nu} K_\nu^\frac{1}{17} F_{\nu_1}^\frac{8}{17} D_L^\frac{16}{17} \nu_1^\frac{p-36}{34} \delta^{-\frac{12+p}{34}} \nonumber \\ & (1+z)^{\frac{p-52}{34}} (1+\eta)^{\frac{1}{17}} \epsilon_\nu^\frac{1}{17} ,
\end{align}

\begin{align}\label{eq:final_B_nu_form}
    B_\text{homog} = & k_1^\frac{1}{17} k_{3\text{,}\nu}^4 K_\nu^\frac{4}{17}F_{\nu_1}^{-\frac{2}{17}} D_L^{-\frac{4}{17}} \nu_1^\frac{2p+13}{17} \delta^{-\frac{2p+7}{17}} \nonumber \\ & (1+z)^{\frac{2p+15}{17}} (1+\eta)^{\frac{4}{17}} \epsilon_\nu^{\frac{4}{17}}, 
\end{align}

\begin{align}
    N_{0,\text{homog}} = & \frac{11}{48\pi} \left( k_1^\frac{1}{17} k_{3\text{,}\nu}^4 F_{\nu_1}^{-\frac{2}{17}} D_L^{-\frac{4}{17}} \nu_1^\frac{2p+13}{17} \delta^{-\frac{2p+7}{17}} \nonumber (1+z)^{\frac{2p+15}{17}} \right)^{\frac{6-p}{2}} \\ & (1+\eta)^{-\frac{5+2p}{17}} \epsilon_\nu^{-\frac{5+2p}{17}} K_\nu^{-\frac{5+2p}{17}} .
\end{align}

\noindent Where we have calculated $N_{0,\text{homog}}$ using equation~\eqref{eq:final_B_nu_form} and \eqref{eq:epsilon_nu}. The density of non-thermal electrons, $n_e$ in the emitting region can be calculated from $N_{0,\text{homog}}$ by considering its definition in equation~\eqref{eq:N0_definition} and equation~\eqref{eq:nu_crit}:

\begin{align}
    n_e = & N_0 B^{\frac{p-1}{2}} \int_{\nu_\text{min}}^{\nu_\text{max}} \frac{1}{2} \nu^\frac{-p-1}{2} c_1^{\frac{1-p}{2}} d\nu \nonumber \\ = & \frac{11}{48\pi} K_{n,\nu} \: k_1^\frac{5}{34} k_{3\text{,}\nu}^{10} K_\nu^{-\frac{7}{17}} F_{\nu_1}^{-\frac{10}{34}} D_L^{-\frac{20}{34}} \nu_1^\frac{10p+65}{34} \nonumber \\ & \delta^{-\frac{10p+35}{34}}  (1+z)^{\frac{10p+75}{34}} (1+\eta)^{-\frac{7}{17}} \epsilon_\nu^{-\frac{7}{17}} , 
\end{align}

\noindent where $K_{n,\nu} = \int_{\nu_\text{min}}^{\nu_\text{max}} \frac{1}{2} \nu^\frac{-p-1}{2} c_1^{\frac{p-1}{2}} d\nu$. Then from $n_e$ the total number of non-thermal electrons is $N_e = \frac{4}{3}\pi R_\text{homog}^3 n_e$ for a quasi-spherical emitting region:

\begin{align}
    N_e = & \frac{11}{36}  K_{n,\nu} \:k_1^{-\frac{19}{34}} k_{3\text{,}\nu}^{13} K_\nu^{-\frac{4}{17}} F_{\nu_1}^{\frac{19}{17}} D_L^{\frac{38}{17}} \nu_1^\frac{13p-43}{34} \delta^{-\frac{13p+71}{34}} \nonumber \\ & (1+z)^{\frac{13p-81}{34}} (1+\eta)^{-\frac{4}{17}} \epsilon_\nu^{-\frac{4}{17}} . 
\end{align}

\noindent We can also derive the intrinsic brightness temperature of the emitting region, defined as:

\begin{equation}
    T_B = \frac{I_\nu c^2 }{2  k_B \nu^2 R^2} =  \frac{c^2 F_{\nu_1} D_L^2}{2\pi k_B \nu_1^2 R^2} \delta^{-1} (1+z)^{-3}.
\end{equation}

\noindent Using the equation for $R_\text{homog}$ this gives the formulae for brightness temperature in terms of observables as:

\begin{align}
    T_B = & \frac{c^2}{2\pi k_B} k_1^\frac{8}{17} k_{3,\nu}^{-2} K_\nu^{-\frac{2}{17}} F_\nu^\frac{1}{17} D_L^{\frac{2}{17}} \nu_1^{\frac{2-p}{17}} \nonumber \\ & \delta^{\frac{p-5}{17}} (1+z)^{\frac{1-p}{17}} (1+\eta)^{-\frac{2}{17}} \epsilon_\nu^{-\frac{2}{17}} ,
\end{align}

\noindent and we see that the brightness temperature is only weakly dependent on the observables, meaning it is likely always in the range $10^{10}-10^{11}~\text{K}$, as noted by \cite{fender_bright_2019}.

Finally, we can use the observation of a spectral index $>\frac{1}{3}$ below the peak of the spectrum to constrain the Lorentz factor of the electrons at the low energy cut-off. If the spectral index is steeper than $\frac{1}{3}$ below the peak then this requires $\frac{\nu_1}{\nu_\text{mins}}>1$, where $\nu_1$ and $\nu_\text{min}$ are in the observer frame. Then:

\begin{equation}
    \nu_\text{min} = c_1B(\gamma_\text{min}m_ec^2)^2\frac{\delta}{(1+z)} , 
\end{equation}

\noindent and substituting this and equation~\eqref{eq:final_B_nu_form} into the inequality gives:

\begin{align}
    \gamma_\text{min} < & \left(c_1^{-1} k_1^{-\frac{1}{17}} k_{3,\nu}^{-4} K_\nu^{-\frac{4}{17}} F_{\nu_1}^{\frac{2}{17}} D_L^{\frac{4}{17}} \nu_1^{\frac{4-2p}{17}} \right. \nonumber \\ & \left.  \delta^{\frac{2p-10}{17}} (1+z)^{\frac{2-2p}{17}} (1+\eta)^{-\frac{4}{17}} \epsilon_\nu^{-\frac{4}{17}} \right)^\frac{1}{2} \frac{1}{m_ec^2}
\end{align}

\noindent Formally, this inequality must be solved numerically, and requires knowledge of the high frequency cut-off (or high energy cut-off in equation~\eqref{eqn:gamma_min_constraint}). However, in practice the dependency on $K_\nu$ (or $K_E$) is so weak that for $p>2$ there is variation of less than $\sim40\%$ in the $\gamma_\text{min}$ case over 10 orders of magnitude in $\nu_\text{max}$ (or $E_\text{max}$).

The above equations for all quantities are in the frequency cut-off formalism. Below we provide the same set of equations in the energy cut-off formalism:

\begin{align}\label{eq:final_E_e_form}
    E_\text{homog} = & \frac{1+2(p+6)}{2(p+1)} \frac{4}{3} \frac{1}{8} k_{3\text{,}E}^{11} k_1^{-\frac{(3p+14)}{2(1+2(p+6))}} K_E^{\frac{11}{1+2(p+6)}} \nu_1^{-1} D_L^{\frac{4+6(p+4)}{1+2(p+6)}}  \nonumber \\ & F_{\nu_1}^{\frac{2+3(p+4)}{1+2(p+6)}} \delta^{-\frac{1+7(p+4)}{1+2(p+6)}} (1+z)^{-\frac{2+5(p+5)}{1+2(p+6)}} (1+\eta)^{\frac{11}{1+2(p+6)}} \nonumber \\ &\left(\frac{2(1+p)}{1+2(p+6)}\epsilon_E^\frac{11}{1+2(p+6)} + \frac{11}{1+2(p+6)}\epsilon_E^{-\frac{2(1+p)}{1+2(p+6)}}\right) ,
\end{align}

\begin{align}\label{eq:final_r_e_form}
    R_\text{homog} = & k_{3\text{,}E} k_1^{-\frac{p+6}{2(1+2(p+6))}} K_E^{\frac{1}{1+2(p+6)}} \nu_1^{-1} D_L^{\frac{2(p+6)}{1+2(p+6)}} F_{\nu_1}^{\frac{p+6}{1+2(p+6)}} \nonumber \\ & \delta^{-\frac{p+5}{1+2(p+6)}} (1+z)^{-\frac{1+3(p+6)}{1+2(p+6)}} (1+\eta)^{\frac{1}{1+2(p+6)}} \epsilon_E^{\frac{1}{1+2(p+6)}},
\end{align}

\begin{align}\label{eq:final_B_e_form}
    B_\text{homog} = & k_{3\text{,}E}^4 k_1^{\frac{1}{1+2(p+6)}} K_E^\frac{4}{1+2(p+6)} \nu_1 D_L^{-\frac{4}{1+2(p+6)}} F_{\nu_1}^{-\frac{2}{1+2(p+6)}} \nonumber \\ & \delta^{-\frac{2p+7}{1+2(p+6)}} (1+z)^{\frac{1+2(p+7)}{1+2(p+6)}} (1+\eta)^{\frac{4}{1+2(p+6)}} \epsilon_E^{\frac{4}{1+2(p+6)}} ,
\end{align}

\begin{align}
    N_{0,\text{homog}} = & \frac{11}{16\pi (p+1)} k_{3,E}^{8} k_1^{\frac{2}{1+2(p+6)}} K_E^{-\frac{5+2p}{1+2(p+6)}} \nu_1^2 \nonumber \\ & D_L^{-\frac{8}{1+2(p+6)}} F_{\nu_1}^{-\frac{4}{1+2(p+6)}} \delta^{-\frac{4p+14}{1+2(p+6)}} (1+z)^{\frac{2+4(p+7)}{1+2(p+6)}} \nonumber \\ & (1+\eta)^{-\frac{5+2p}{1+2(p+6)}} \epsilon_E^{-\frac{5+2p}{1+2(p+6)}} , 
\end{align}

\begin{equation}
    n_e =  N_{0,\text{homog}} \int^{E_\text{max}}_{E_\text{min}} E^{-p} dE = N_{0,\text{homog}} \: K_{n,E} \: , 
\end{equation}

\begin{align}
    N_e = & \frac{11}{12(p+1)} K_{n,E} \: k_{3,E}^{11} k_1^{-\frac{14+3p}{2(1+2(p+6))}} K_E^{-\frac{2+2p}{1+2(p+6)}} \nu_1^{-1} \nonumber \\ & D_L^{\frac{4+6(p+4)}{1+2(p+6)}} F_{\nu_1}^\frac{2+3(p+4)}{1+2(p+6)}  \delta^{-\frac{1+7(p+4)}{1+2(p+6)}} (1+z)^{-\frac{2+5(p+5)}{1+2(p+6)}} \nonumber \\ & (1+\eta)^{-\frac{2+2p}{1+2(p+6)}} \epsilon_E^{-\frac{2+2p}{1+2(p+6)}} ,
\end{align}

\begin{align}
    T_B = & k_{3,E}^{-2} k_1^{\frac{p+6}{1+2(p+6)}} K_E^{-\frac{2}{1+2(p+6)}} D_L^{\frac{2}{1+2(p+6)}} F_{\nu_1}^{\frac{1}{1+2(p+6)}} \delta^{-\frac{3}{1+2(p+6)}} \nonumber \\ & (1+z)^{-\frac{1}{1+2(p+6)}} (1+\eta)^{-\frac{2}{1+2(p+6)}} \epsilon_E^{-\frac{2}{1+2(p+6)}} ,
\end{align}

\begin{align}\label{eqn:gamma_min_constraint}
    \gamma_\text{min} < & \left(c_1^{-1} k_{3,E}^{-4} k_1^{-\frac{1}{1+2(p+6)}} K_E^{-\frac{4}{1+2(p+6)}} D_L^\frac{4}{1+2(p+6)} F_{\nu_1}^\frac{2}{1+2(p+6)} \delta^{-\frac{6}{1+2(p+6)}} \right. \nonumber \\ & \left. (1+z)^{-\frac{2}{1+2(p+6)}} (1+\eta)^{-\frac{4}{1+2(p+6)}} \epsilon_E^{-\frac{4}{1+2(p+6)}} \right)^{\frac{1}{2}} \frac{1}{m_ec^2} .
\end{align}

\noindent Finally, we add the caution that the important observable parameters for all of these equations are the frequency of the emitting region where the optical depth is unity, $\nu_1$, and the associated flux density, $F_{\nu_1}$. These are related to the peak in the synchrotron spectrum by equation~\eqref{taumax}. However, as previously stated in Section~\ref{sec:trad_estimates}, and as shown in Section~\ref{sec:time}, accurately measuring $\nu_1$ and $F_{\nu_1}$ from a lightcurve can be more nuanced.

\section{The energy cut-off formalism}\label{sec:energy_cutoff_form}

As demonstrated in Section~\ref{sec:trad_estimates} the form of the traditional homogeneous SSA equations to calculate the minimum energy and associated parameters depends on whether knowledge of frequency cut-offs in the synchrotron spectrum or energy cut-offs in the electron spectrum is assumed. The difference in the form of the equations arises as the frequency cut-offs encode information about both the energy cut-offs and the magnetic field (see equation~\eqref{eq:nu_crit}). In Section~\ref{sec:trad_estimates} we derived the equations in the frequency cut-off formalism, using $E_e = V N_0 B^{\frac{p-2}{2}}K_\nu$. Here we instead derive the energy cut-off formalism, using $E_e = V N_0 K_E$ where:

\begin{equation}
    K_E = \int^{E_\text{max}}_{E_\text{min}}E^{1-p} dE ,
\end{equation}

\noindent and follow a similar process as in Section~\ref{sec:trad_estimates}. As before, we have $V\propto R^3$, $B \propto R^4$ and $N_0\propto R^{-(2p+5)}$. Then the total energy can be written as:

\begin{equation}
    E_\text{total} = E_B+E_e \propto R^{11} + R^{-2(p+1)} .
\end{equation}

\noindent The total energy is then minimised with respect to $R$ when:

\begin{equation}
    E_B = \frac{2(1+p)}{11} E_e .
\end{equation}

\noindent Using this combined with equations~\eqref{nu1_eqn} and \eqref{fnu1_eqn}, and quasi-sphericity, we can solve for the minimum energy and associated parameters:

\begin{align}\label{eq:e_homog_E}
    E_\text{homog, min} = & \frac{1+2(p+6)}{2(p+1)} \frac{4}{3} \frac{1}{8} k_{3\text{,}E}^{11} k_1^{-\frac{(3p+14)}{2(1+2(p+6))}} K_E^{\frac{11}{1+2(p+6)}} \nonumber \\ &\nu_1^{-1} D^{\frac{4+6(p+4)}{1+2(p+6)}} F_{\nu_1}^{\frac{2+3(p+4)}{1+2(p+6)}} .
\end{align}

\begin{equation}\label{eq:r_homog_E}
    R_\text{homog, eq} = k_{3\text{,}E} k_1^{-\frac{2(p+6)}{4(1+2(p+6))}} K_E^{\frac{1}{1+2(p+6)}} \nu_1^{-1} D^{\frac{2(p+6)}{1+2(p+6)}} F_{\nu_1}^{\frac{p+6}{1+2(p+6)}} ,
\end{equation}

\begin{equation}\label{B_homog_E}
    B_\text{homog, eq} = k_{3\text{,}E}^4 k_1^{\frac{1}{1+2(p+6)}} K_E^\frac{4}{1+2(p+6)} \nu_1 D^{-\frac{4}{1+2(p+6)}} F_{\nu_1}^{-\frac{2}{1+2(p+6)}} ,
\end{equation}

\noindent where $k_{3,E}$ is a constant weakly dependent on $p$ defined as:

\begin{equation}
    k_{3,E}(p) =\left( \frac{11}{2(p+1)} \frac{1}{8 \pi} \frac{4}{3} c_6(p)\right)^{-\frac{1}{1+2(p+6)}} (2c_1)^{-\frac{p+4}{2+4(p+6)}} .
\end{equation}

\noindent The equations for the minimum energy in the energy cut-off formalism differ from slightly the frequency cut-off formalism presented in Section~\ref{sec:trad_estimates}. However, in the special case of $p=2$ then $k_{3,E}=k_{3\text{,}\nu}$, $K_E=K_\nu$ and both formalisms become equivalent. This is because when $p=2$ the energy in non-thermal electrons can be determined using the frequency cut-offs without needing knowledge of the magnetic field (see equation~\eqref{eq:e_energy_nu}).

\section{Non-thermal protons and the effect of deviation from equipartition}\label{sec:equipartition_deviation}

If there is a significant fraction of energy in non-thermal protons then the minimum energy of the emitting region occurs when the total energy in non-thermal particles (including protons) and the magnetic field energy are close to equipartition. If non-thermal protons are present then the total energy of the emitting region can be written as:

\begin{equation}
    E_\text{tot} = E_B +E_e+E_p = E_B + (1+\eta)E_e , 
\end{equation}

\noindent where $\eta = \frac{E_p}{E_e}$ is defined as measure of the amount of energy in non-thermal protons compared to non-thermal electrons, following e.g. \cite{cowie_2026}. Then following the same procedure as in Section~\ref{sec:trad_estimates} we find that the minimum energy for unknown size occurs when:

\begin{equation}\label{ratio_protons_nuform}
    E_B = (1+\eta)\frac{6}{11}E_e , 
\end{equation}

\noindent in the frequency cut-off formalism or 

\begin{equation}\label{ratio_protons_eform}
    E_B = (1+\eta)\frac{2(1+p)}{11}E_e , 
\end{equation}

\noindent in the energy cut-off formalism. Continuing the derivation in Section~\ref{sec:trad_estimates} using the above equations including non-thermal protons results in the minimum energy and other quantities having a dependence on $(1+\eta)$ of:

\begin{equation}
    E_\text{homog, min} \propto (1+\eta)^{\frac{11}{1+2(p+6)}} ,
\end{equation}

\begin{equation}
    B_\text{homog, eq} \propto (1+\eta)^{\frac{4}{1+2(p+6)}} ,
\end{equation}

\begin{equation}
    R_\text{homog, eq} \propto (1+\eta)^{\frac{1}{1+2(p+6)}} ,
\end{equation}

\noindent in the energy cut-off formalism. In the frequency cut-off formalism the exponents change from $\frac{X}{1+2(p+6)} \rightarrow \frac{X}{17}$. In general the emitting region may not be close to equipartition and then the energy in non-thermal particles and in the magnetic field will not obey equations~\eqref{ratio_protons_nuform} or \eqref{ratio_protons_eform}. It is useful to parametrise how deviations from equipartition effect the energy and other derived quantities to quantify uncertainties or in case the deviation from equipartition is known \textit{a priori}. We can define the deviation from equipartition using the quantity $\epsilon_\nu$ in the cut-off frequency formalism or $\epsilon_E$ in the cut-off energy formalism defined as:

\begin{equation}\label{eq:epsilon_nu}
    \epsilon_\nu = \frac{E_B}{E_e} \frac{11}{6(1+\eta)} ,
\end{equation}

\noindent and

\begin{equation}\label{eq:epsilon_E}
    \epsilon_E = \frac{E_B}{E_e} \frac{11}{(2p+1)(1+\eta)} ,
\end{equation}

\noindent  respectively. When $\epsilon_\nu$ or $\epsilon_E$ is unity then the energy of the emitting region is the minimum energy, and the non-thermal particles and the magnetic field in the emitting region are close to equipartition. Using this equation in place of equation~\eqref{eq_min_eng_condition} in Section~\ref{sec:trad_estimates} results in:

 \begin{equation}\label{r_scaling_epsilon_nu}
     R_\text{homog} \propto \epsilon_\nu^{\frac{1}{17}} , 
 \end{equation}

\noindent or
 
\begin{equation}\label{r_scaling_epsilon_E}
     R_\text{homog} \propto \epsilon_E^{\frac{1}{1+2(p+6)}} , 
\end{equation}

\noindent depending on the formalism used. Then using the above, equation~\eqref{eq:enrgy_dependence_size}, and equations~\eqref{eq:epsilon_nu} and \eqref{eq:epsilon_E} we find that the energy calculated (which is no longer a minimum if $\epsilon_\nu$ or $\epsilon_E$ is known) has the dependence:

 \begin{equation}
     E_\text{homog} \propto \left(\frac{6}{17}\epsilon_\nu^\frac{11}{17} + \frac{11}{17}\epsilon_\nu^{-\frac{6}{17}}\right) , 
 \end{equation}

\noindent or

 \begin{equation}
     E_\text{homog} \propto \left(\frac{2(1+p)}{1+2(p+6)}\epsilon_E^\frac{11}{1+2(p+6)} + \frac{11}{1+2(p+6)}\epsilon_E^{-\frac{2(1+p)}{1+2(p+6)}}\right)
 \end{equation}

\noindent depending on the formalism used. The dependence of the magnetic field and $N_0$ of the emitting region on deviation from equipartition is apparent given their dependence on $R_\text{homog}$ (discussed in Section~\ref{sec:trad_estimates}) and the equations~\eqref{r_scaling_epsilon_nu} and \eqref{r_scaling_epsilon_E}. These dependencies are shown explicitly in Section~\ref{sec:full_equations}.
 
\section{Corrections for Doppler boosting and redshift}\label{sec:doppler_and_redshift}

If the emitting region has relativistic bulk motion and/or is located at a redshift, $z$, then the equations for the minimum energy estimate and associated parameters must be adjusted. The observed frequency where the optical depth of the emitting region is unity, $\nu^{'}_{1}$, is related to the intrinsic SSA peak frequency by:

\begin{equation}\label{frequency_doppler_redshift}
    \nu^{'}_{1} = \nu_1 \frac{\delta}{(1+z)} , 
\end{equation}

\noindent where $\delta$ is the Doppler factor of the emitting region defined as:

\begin{equation}
    \delta = \frac{1}{\Gamma (1-\beta\cos\theta)} , 
\end{equation}

\noindent where $\beta$ is the bulk velocity in units of the $c$ of the emitting region, $\theta$ is the angle of the bulk velocity vector to the line of sight of the observer, and $\Gamma$ is the bulk Lorentz factor of the emitting region. The observed specific intensity of the synchrotron radiation at $\nu^{'}_{1}$, $I^{'}_{\nu^{'}_{1}}$ is related to the intrinsic specific intensity at $\nu_1$ by:

\begin{equation}\label{intensity_transform}
    I^{'}_{\nu^{'}_{1}} = I_{\nu_1} \frac{\delta^{3}}{(1+z)^3} .
\end{equation}

\noindent Then using equation~\eqref{flux_eq} (valid if the emission is isotropic in the rest frame) we obtain the relationship between the observed flux density at an optical depth of unity, $F^{'}_{\nu^{'}_{1}}$, and the intrinsic flux density at an optical depth of unity:

\begin{equation}\label{flux_transform}
   F^{'}_{\nu^{'}_1} = \frac{\pi R^2}{D^2_A} \frac{\delta^3}{(1+z)^3} I_{\nu_1} = \frac{\pi R^2}{D^2_L} \delta^3 (1+z) I_{\nu_1} , 
\end{equation}

\noindent where we have used that the angular diameter distance, $D_A$ is related to observed solid angle by $\Omega = \frac{\pi R^2}{D^2_A}$, and that the luminosity distance, $D_L$ is related to the angular diameter distance by $D_L = (1+z)^2 D_A$ \citep{hogg_1999}. Here we have assumed that the emitting region is a discrete region as opposed to a continuous jet (see e.g. \citealt{urry_1995} Section 9B for further discussion). Using equation~\eqref{flux_transform} in place of equation~\eqref{flux_eq} in the derivation presented in Section~\ref{sec:trad_estimates} results in the following dependencies on doppler factor and redshift for the minimum energy and related quantities:

\begin{equation}
    E_\text{homog, min} \propto (1+z)^\frac{11p-96}{34} \delta^{-\frac{64+11p}{34}} ,
\end{equation}

\begin{equation}
    R_\text{homog, eq} \propto (1+z)^\frac{p-52}{34} \delta^{-\frac{12+p}{34}} , 
\end{equation}

\begin{equation}
    B_\text{homog, eq} \propto (1+z)^\frac{2p+15}{17} \delta^{-\frac{2p+7}{17}} , 
\end{equation}

\noindent in the frequency cut-off formalism, and

\begin{equation}
    E_\text{homog, min} \propto (1+z)^{-\frac{2+5(p+5)}{1+2(p+6)}} \delta^{-\frac{1+7(p+4)}{1+2(p+6)}} ,
\end{equation} 

\begin{equation}
    R_\text{homog, eq} \propto (1+z)^{-\frac{1+3(p+6)}{1+2(p+6)}} \delta^{-\frac{p+5}{1+2(p+6)}} , 
\end{equation}

\begin{equation}
    B_\text{homog, eq} \propto (1+z)^\frac{1+2(p+7)}{1+2(p+6)} \delta^{-\frac{2p+7}{1+2(p+6)}} , 
\end{equation}

\noindent The full equations including these dependencies are given in Section~\ref{sec:trad_estimates}. The scaling of the different quantities with redshift and Doppler factor agrees with that presented in \cite{matsumoto_2023} (when setting $p=2$, which \cite{matsumoto_2023} implicitly assume) and differs only from that presented in \cite{fender_bright_2019} due to their use of the "single frequency approximation" discussed in Section~\ref{sec:comparison_to_prev}. We note that due to the fact that because no matter the Doppler factor or redshift the same part of the synchrotron spectrum is observed, e.g. close to the SSA peak, that the spectral index of has no impact on the boosted observed flux density. In other words we are considering the transform between $F^{'}_{\nu^{'}} \leftrightarrow F_\nu$ rather than $F^{'}_{\nu^{'}} \leftrightarrow F_{\nu^{'}}$ 

\vspace{170pt}

\section{Magnetic field correction factors}\label{sec:mag_correction_appendix}

\begin{center}
 \includegraphics[width=0.9\columnwidth]{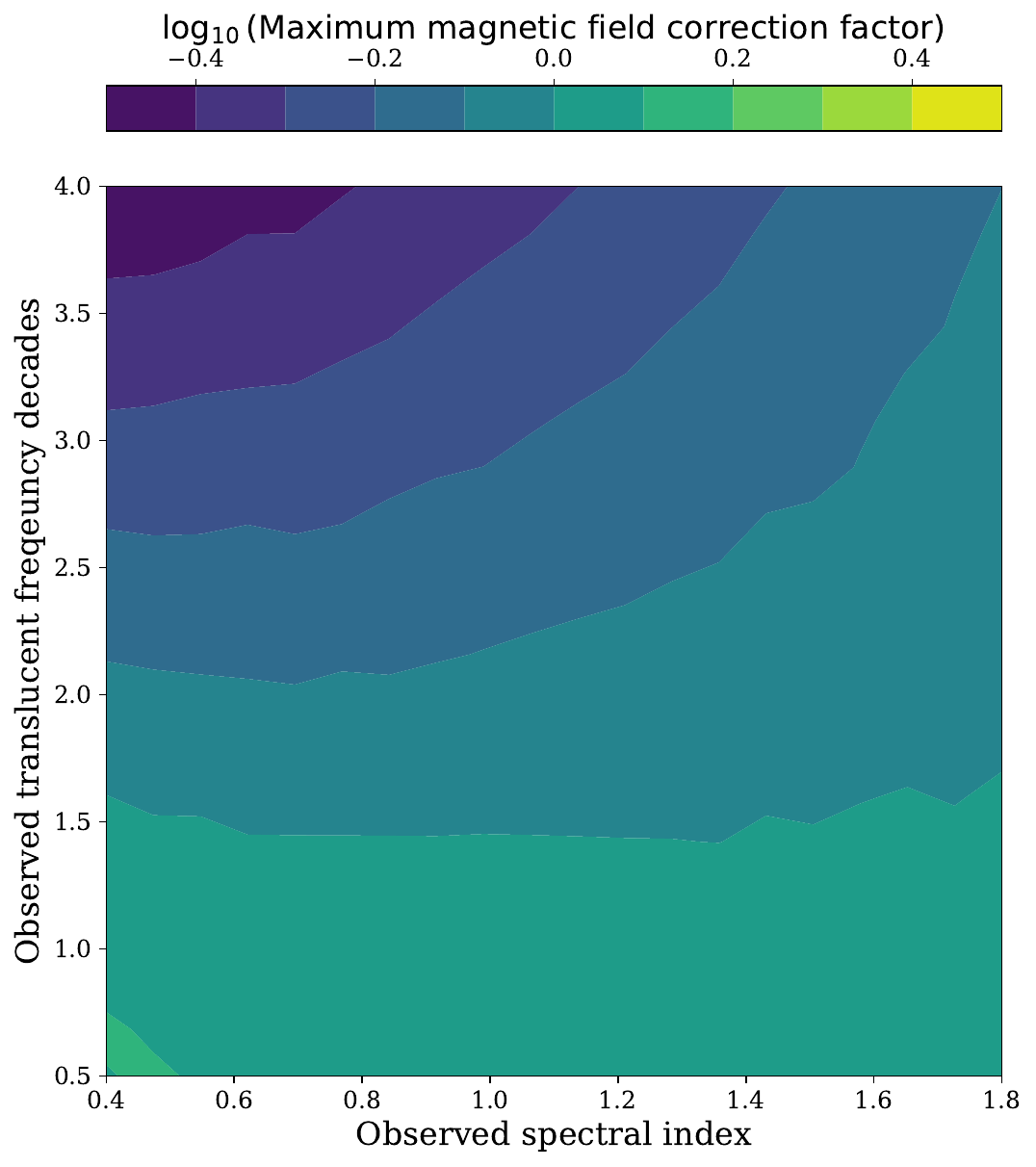}
 \captionof{figure}{A contour plot of the logarithm to base 10 of the maximum magnetic field correction factor between the inhomogeneous slab model with the given observational parameters, and the equivalent homogeneous model.}
 \label{fig:max_B_ratios_contour}
\end{center}

\begin{center}
 \includegraphics[width=0.9\columnwidth]{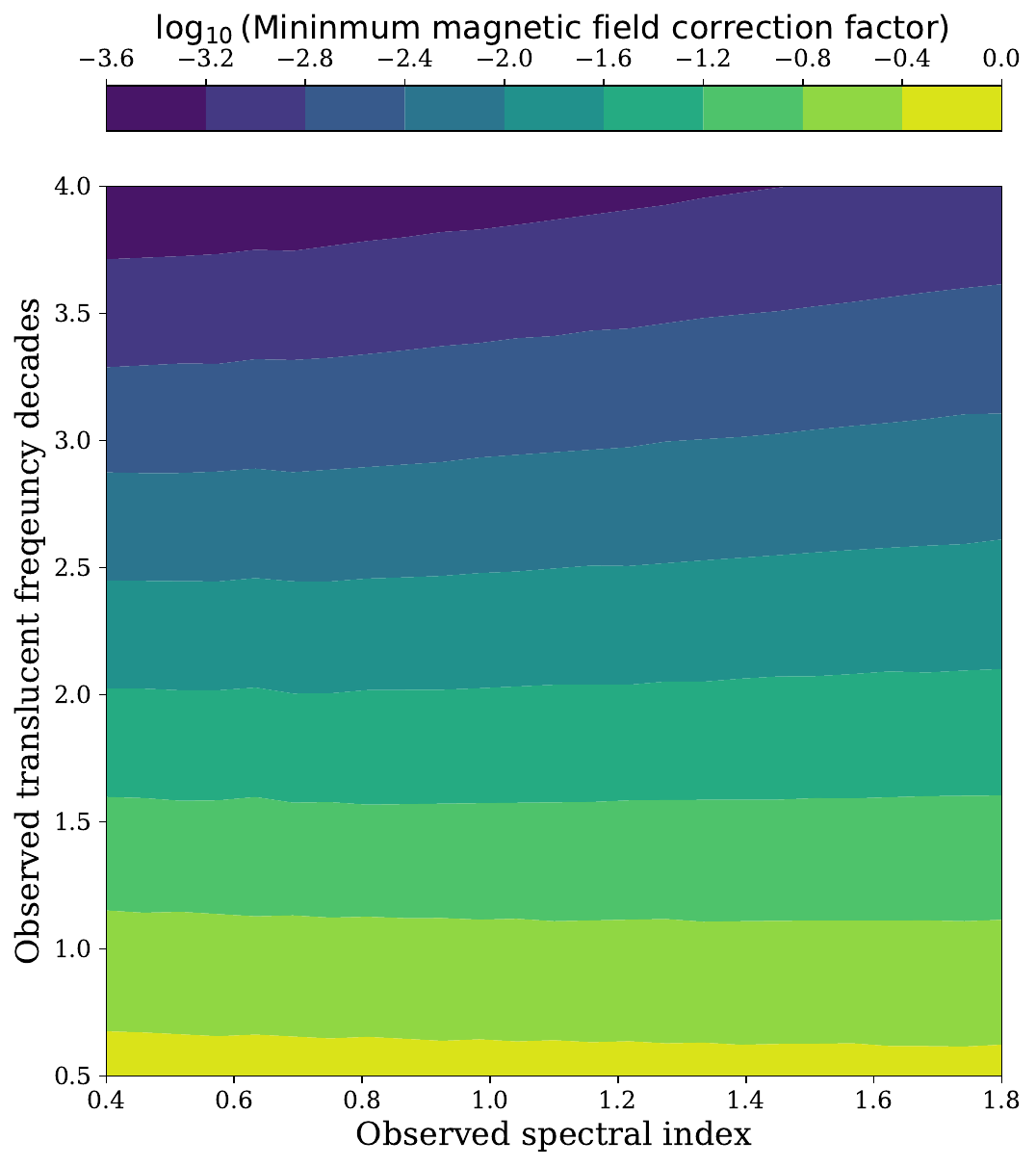}
 \captionof{figure}{A contour plot of the logarithm to base 10 of the minimum magnetic field correction factor between the inhomogeneous slab model with the given observational parameters, and the equivalent homogeneous model.}
 \label{fig:min_B_ratios_contour}
\end{center}


\bsp	
\label{lastpage}
\end{document}